\documentclass[twocolumn,10pt]{IEEEtran}
\usepackage{braket}

\input{def.tex}
\usepackage{dsfont}
\usepackage{minibox}
\DeclareSymbolFont{matha}{OML}{txmi}{m}{it}
\DeclareMathSymbol{\varv}{\mathord}{matha}{118}
\usepackage{eurosym}
\usepackage{hhline,physics}

\usepackage{multicol}
\usepackage{algorithm,algorithmicx}
\usepackage{graphicx}
\usepackage{textcomp}
\usepackage{caption,pifont}
\usepackage[multiple]{footmisc}
\renewcommand{\baselinestretch}{0.99}

\usepackage[english]{babel}

\IEEEoverridecommandlockouts
\begin{document}
	\title{{\LARGE Joint Spatial Division and Multiplexing for FDD in Intelligent Reflecting Surface-assisted Massive MIMO Systems} }
	\author{{\normalsize Anastasios Papazafeiropoulos,  Pandelis Kourtessis, Konstantinos Ntontin, Symeon Chatzinotas} \thanks{A. Papazafeiropoulos is with the Communications and Intelligent Systems Research Group, University of Hertfordshire, Hatfield AL10 9AB, U. K., and with SnT at the University of Luxembourg, Luxembourg. P. Kourtessis is with the Communications and Intelligent Systems Research Group, University of Hertfordshire, Hatfield AL10 9AB, U. K. K. Ntontin and S. Chatzinotas are with the SnT at the University of Luxembourg, Luxembourg. E-mails: tapapazaf@gmail.com,  p.kourtessis@herts.ac.uk, \{kostantinos.ntontin,symeon.chatzinotas\}@uni.lu.}}
	\maketitle
	\begin{abstract}
		Intelligent reflecting surface (IRS) is a promising technology to deliver the higher spectral and energy requirements   in fifth-generation (5G) and beyond wireless networks while shaping the propagation environment. Such a design can be  further enhanced with massive multiple-input-multiple-output (mMIMO) characteristics towards  boosting the network performance.  However, channel reciprocity,  assumed in 5G systems such as mMIMO, appears to be questioned in practice by recent studies on IRS. Hence, contrary to previous works, we consider  frequency division duplexing (FDD)    to study the performance of an IRS-assisted mMIMO system. However,  FDD is not suitable for large number of antennas architectures. For this reason we employ the joint spatial division and multiplexing (JSDM)  approach exploiting the structure of the correlation of the channel vectors to reduce the  channel state information (CSI) uplink feedback, and thus, allowing the use even of a large number of antennas at the base station. JSDM entails dual-structured precoding and clustering  the user equipments (UEs) with the same covariance matrix into groups. Specifically, we derive the sum spectral efficiency (SE) based on statistical CSI in terms of large-scale statistics by using the deterministic equivalent (DE) analysis while accounting for correlated Rayleigh fading. Subsequently, we formulate the optimization problem concerning the sum SE with respect to the reflecting  beamforming matrix (RBM) and the total transmit power, which can be performed at every several coherence intervals by taking advantage  of the slow-time variation of the large-scale statistics. This notable property contributes further to the decrease of the feedback overhead. Numerical results, verified by Monte-Carlo (MC) simulations,  enable interesting observations by  elucidating how fundamental system parameters such as the rank of the covariance matrix and the number of groups of UEs affect the performance. For example, the selection of a high  rank improves the channel conditioning but increases the feedback overhead.
	\end{abstract}
	
	\section{Introduction}
	The stringent requirements of fifth-generation  (5G)  and beyond wireless networks in terms of spectral efficiency (SE) and energy efficiency (EE) could be achieved by relying on disruptive technologies such as intelligent reflecting surface (IRS) and massive multiple-input multiple-output (mMIMO) systems  \cite{Zhang2020b,Marzetta2016,Wu2020}. In particular, IRS has been recently emerged as a revolutionary solution supporting high data rates and low energy consumption \cite{Wu2019,Huang2019,Pan2020,Guo2020,Han2019,Elbir2020,Yang2020,Wu2020a,Tang2021a}. It consists of nearly passive elements, which can create favourable propagation conditions, especially against obstacles, by a smart adjustment of the elements' phase shifts on the impinging waves. Besides IRS, the mMIMO technology, which has been already deployed in 5G systems \cite{Butovitsch2018}, assumes a large number of antennas at each base station (BS) by applying low-complexity linear precoding techniques. Among its benefits, we meet high directional beamforming and multiplexing gains while almost cancelling intra-cell interference in the large antenna regime due to channel orthogonality.

	With respect to IRS, previous works are classified  in two categories depending on whether the phase shifts optimization relies on the instantaneous channel state information (I-CSI) \cite{Wu2019,Huang2019,Pan2020} or statistical CSI (S-CSI) \cite{Guo2020,Han2019,Peng2020,Yu2020,Hu2020,Nadeem2021,Kammoun2020,Zhao2020,Papazafeiropoulos2021,VanChien2021,Papazafeiropoulos2021a}.  For example, in the first category,  requiring optimization at each coherence interval, active and passive beamforming were jointly considered  for single cell multi-antenna systems  in \cite{Wu2019} while the EE maximization problem was addressed in \cite{Huang2019}.
	
	In the other category, phase shifts, optimized based on S-CSI, do not need to change at every coherence interval but every several intervals. Compared to I-CSI, this method offers lower feedback overhead due to less frequent phase optimization, especially when the numbers of BS antennas and IRS elements are large. Moreover, the feedback overhead reduction  results in less power consumption by the IRS controller and release of the capacity requirement for the IRS control link. Furthermore, a significant computational complexity reduction is achieved at the BS, which needs to update the phase shift matrix at a much larger time scale compared to phase shift matrix update in I-CSI schemes that occurs at a much lower frequency.  As a result, I-CSI designs could be applied to scenarios with low mobility or fixed location while, in the cases of short coherence intervals such as high mobility environments, phase shifts tuning  based on S-CSI tends to be an advantageous solution.  On the  ground of these benefits, S-CSI design has attracted a lot of interest \cite{Guo2020,Han2019,Peng2020,Yu2020,Hu2020,Nadeem2021,Kammoun2020,Zhao2020,Papazafeiropoulos2021,VanChien2021,Papazafeiropoulos2021a}.  For instance, in \cite{Han2019}, optimal IRS  phase shift design was presented based on maximization of the ergodic rate. Further, in \cite{Peng2020}, the authors studied IRS-assisted multi-pair  systems and applied a genetic algorithm during the phase shifts optimization. 
	Taking into account for random matrix theory, the authors in \cite{Kammoun2020}  studied the minimum signal-to-interference-plus-noise ratio (SINR) maximization. In \cite{Zhao2020}, a novel two-timescale beamforming
	optimization scheme was proposed, where first S-CSI was used to optimize the passive beamforming, and then I-CSI was used to design the active beamforming. In \cite{Papazafeiropoulos2021}, both IRS and additive transceiver hardware impairments (HIs) were studied. Note that IRS HIs appear due to the lack of infinite precision at the IRS phase shifts \cite{Badiu2019}. In \cite{Papazafeiropoulos2021a}, two insightful scenarios, namely, a finite number of large IRS and a large number of finite size IRS were considered towards deriving the coverage probability and it was shown that the latter implementation method is more advantageous.


	Although most works have relied on a time division duplex (TDD) design, recent results regarding IRS showed that the phase shifts depend on the angle of the impinging electromagnetic waves, which renders the assumption of channel reciprocity questionable \cite{Chen2020,Pan2021,Tang2021}. In particular, in \cite{Chen2020}, it was shown that channel reciprocity holds only when the incident angles are small while, in \cite{Tang2021}, several IRS-assisted approaches such as the time-varying control and the structural asymmetries, realizing non-reciprocal channels, were introduced.  Hence, the study of the design and the  performance of   IRS-assisted systems, operating in frequency division 	duplexing (FDD), is  required and is the motivation for the work presented in this paper.\footnote{This observation has been also recognised in \cite{Pan2021} as one of the most interesting topics for future research.}  Note that many current wireless networks are based on FDD systems, which contributes further to our  interest in such systems. Moreover, FDD operates more effectively   in the cases of symmetric traffic and 		 delay-sensitive applications \cite{Jiang2015}.  However, FDD cannot be easily deployed in 5G and beyond systems such as mMIMO because the  multi-antenna channel acquisition makes their implementation prohibitive \cite{Jindal2006}. In particular, FDD includes downlink training, which together with the CSI feedback, present a significant bottleneck as the number of BS antennas increases.  Fortunately, the use  of the  joint spatial division and 	multiplexing (JSDM) approach allows achieving  SE gains similar to mMIMO \cite{Adhikary2013} while it has not been applied yet in IRS systems. 
	
	The concept of JSDM relies on the clustering of UEs  into groups having  approximately the same covariance matrix and the application of dual structured precoding to reduce the feedback overhead efficiently. The structure of the precoder consists of a prebeamforming matrix minimizing the inter-group interference and a linear precoding that depend on large and small-scale fading, respectively. Specifically,  the large-scale fading is slowly varying  because it depends on the channel second-order statistics such as the channel covariance matrices, which can be acquired accurately with a low feedback overhead while the small-scale fading involves instantaneous CSI.  It is worthwhile to mention that the application of JSDM is particularly	 justified in IRS-assisted systems since a large number of IRS elements results in large dimensional channel matrices that incur further high feedback overhead in FDD IRS systems.

	\subsection{Contributions }
	The main contributions are summarized as follows.
	\begin{itemize}
					\item Contrary to existing works such as \cite{Peng2020,Yu2020,Hu2020,Nadeem2021,Kammoun2020,Zhao2020,Papazafeiropoulos2021,VanChien2021,Papazafeiropoulos2021a}, which relied on time division duplexing (TDD) transmission for IRS-assisted  systems, we consider  FDD. To the best of our knowledge, there are no prior works accounting for  FDD in IRS and studying its performance except of \cite{Chen2020a,Guo2021,Shen2021}.  Especially, \cite{Chen2020a} proposed a cascaded codebook and an adaptive bit partitioning strategy, \cite{Guo2021} considered a  two-way passive beamforming design, and \cite{Shen2021} proposed a dimension reduced channel feedback scheme  to reduce the channel feedback overhead in RIS assisted systems.
		\item This is the first work applying the concept of JSDM \cite{Adhikary2013} on IRS-assisted systems to  reduce the feedback overhead being increasingly significant with the numbers of antennas and IRS elements. In particular, not only do we study the performance but we also provide system guidelines concerning the choice of JSDM parameters. Moreover, we have generalized \cite{Adhikary2013} since we do not assume equal power  among the UEs but perform power optimization.
		\item Although many previous works were based on I-CSI knowledge and independent Rayleigh fading such as \cite{Wu2019,Huang2019,Pan2020}, we move towards consideration of realistic conditions and we hinge on S-CSI (more advantageous) and correlated Rayleigh fading, which is unavoidable as shown in \cite{Bjoernson2020}. Remarkably, based on the use of S-CSI and the effective channel, consisting of the cascaded and direct channels, we achieve to obtain the achievable SE of IRS-assisted systems similar to conventional systems. Notably, works considering S-CSI and correlated Rayleigh fading exist  (e.g., \cite{Kammoun2020,Zhao2020,Papazafeiropoulos2021,Papazafeiropoulos2021a}) but none of them has accounted for the  JSDM approach.
			\item By employing the per-group processing  (PGP) method of the JSDM approach to reduce the feedback overhead \cite{Adhikary2013}, we derive the sum SE of the IRS-assisted mMIMO  systems based on  FDD in closed-form by leveraging results from the deterministic equivalent (DE) analysis \cite{Wagner2012,Hoydis2013,Papazafeiropoulos2015a}. The importance of the results in terms of DEs is noteworthy since DEs provide an analytical tool  resulting in a deterministic expression of the SE based on a convergent system of fixed-point 	equations while avoiding lengthy Monte Carlo simulations.
		\item		We formulate the optimization problem concerning the sum SE subject to reflecting beamforming matrix (RBM)  and total transmit power constraints. Note that the PGP method includes the regularized zero-forcing (RZF)  precoding and this is the unique work providing RBM  optimization based on S-CSI with such a complex precoder. Other works such as \cite{Kammoun2020}  considered a simple linear precoding, e.g., maximum ratio transmission (MRT) precoding.		Furthermore, the proposed optimization is based on deterministic expressions dependent only on large-scale statistics, and thus, can be performed at every several coherence intervals and reduce considerably the signal overhead as required especially in FDD systems.				
		\item We verify the results by Monte-Carlo (MC) simulations and we shed light on the impact of the system parameters on the sum SE such as the signal-to-noise ratio (SNR), the effective covariance rank, the effective channel dimension, and the number of IRS elements. Especially, the effective covariance rank and effective channel dimension play a prominent role by presenting a trade-off between the feedback reduction and the performance.
	\end{itemize}
	
	\subsection{Paper Outline} 
	The remainder of this paper is organized as follows. Section~\ref{System} presents the system model of an IRS-assisted massive MIMO system with correlated Rayleigh fading operating in FDD and employing the JSDM. Section~\ref{AsymptoticPerformance} provides the asymptotic performance analysis. Section~\ref{SumSEMaximizationDesign} presents the sum SE maximization with respect to  the IRS RBM and the transmit power.
	The numerical results are placed in Section~\ref{Numerical}, and Section~\ref{Conclusion} concludes the paper.
	
	\subsection{Notation}Vectors and matrices are denoted by boldface lower and upper case symbols, respectively. The notations $(\cdot)^\T$, $(\cdot)^\H$, and $\tr\!\left( {\cdot} \right)$ represent the transpose, Hermitian transpose, and trace operators, respectively. The expectation operator is denoted by $\EE\left[\cdot\right]$  while $ \diag\left(\ba \right) $ represents an $ n\times n $ diagonal matrix with diagonal elements being the elements of vector $ \ba $. 
		In the case of a matrix $ \bA $, $\diag\left(\bA\right) $ denotes  a vector with elements the diagonal elements of $ \bA $.
	Also, we denote $ [x]^{+}=\max(0,x) $ and the notations $\xrightarrow[ M \rightarrow \infty]{\mbox{a.s.}}$ and $a_n\asymp b_n$ with $a_n$ and $b_n$ being two infinite sequences denote almost sure convergence as $ M \rightarrow \infty $. The notations $ \mathrm{Span}(\bX) $ and $\mathrm{Span}^{\bot}(\bX)  $ denote the column space of $ \bX $ and 	its orthogonal complement, respectively. Also, the notation $ \bX_{[k]} $ denotes the matrix obtained upon removing the $ k $th column of $ \bX $, while $\bb \sim \cC\cN{(\b0,\mathbf{\Sigma})}$ represents a circularly symmetric complex Gaussian vector with {zero mean} and covariance matrix $\mathbf{\Sigma}$. Finally, Table \ref{AbbsTable} provides the main abbreviations for the sake of convenience. \begin{table}
	\begin{center}
				\begin{tabular}{ | c | c |  }
					\hline
					Intelligent reflecting surface  & IRS  \\ \hline
					Frequency division duplexing  &FDD \\ \hline
					Time division duplexing  &TDD \\ \hline
					Joint spatial division and multiplexing & JSDM \\
					\hline
					Statistical CSI  & S-CSI \\\hline
					Instantaneous channel state information  & I-CSI \\\hline
					Joint spatial division and multiplexing & JSDM \\\hline
					Per-group processing  & PGP \\\hline
					Joint-group processing  & JGP \\\hline
					Reflecting beamforming matrix & RBM \\\hline
					Regularized zero-forcing & RZF \\
					\hline
				\end{tabular}
			\end{center}	
			\caption{  Table of main abbreviations}\label{AbbsTable}
	\end{table}
	\section{System Model}\label{System}
\begin{figure}[!h]
		\begin{center}
			\includegraphics[width=0.85\linewidth]{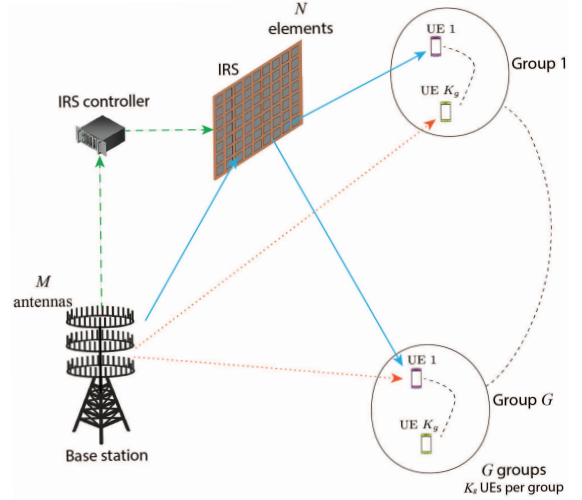}
			\caption{An IRS-assisted downlink MU-MISO communication system with $ M $ BS antennas, $ N $ IRS elements, and $ K $ UEs clustered into $ G $ groups.  }
			\label{Fig1}
		\end{center}
	\end{figure}

	We consider the downlink of an IRS-aided multi-user (MU) multiple-input, single-output (MISO) system with one BS deployed with $ M $ antennas that  serve $K $ single-antenna UEs as shown in Fig. \ref{Fig1}. Additionally to possibly existing direct links between the BS and the UEs, one IRS, implemented with $ N $ nearly passive reflecting elements introducing phase shifts onto the impinging signal waves, assists the communication. The size of each IRS element is $ d_{\mathrm{H}}\times d_{\mathrm{V}} $ with $d_\mathrm{V}$ and $d_\mathrm{H}$ expressing its vertical height and its horizontal width, respectively. The management of the phase shifts takes place by a controller that communicates with the BS through a perfect backhaul link.

	As a reasonable design, we assume that the IRS has a line-of-sight (LoS) with the BS. This assumption can be justified 
	by assuming that  both the BS and IRS are deployed at high altitude and their locations are fixed. Moreover, we assume that the UEs are spatially clustered into $ G $ groups each having $ K_{g} $ UEs with the same spatial covariance matrix. The UEs in the same group are nearly co-located while different groups are well-separated. Note that this is a practical assumption since UEs tend to confine in small regions such as buildings.  Hence, there exist $ K=\sum_{g=1}^{G} K_{g} $ UEs in total, where the index $ g $ refers to  UEs in group $ g $.
	
	\subsection{Channel and Signal Models}\label{ChannelModel} 
	Assuming a flat-fading channel, the received $ K_{g}\times 1 $ complex baseband signal by the UEs at the $ g $th group is given by
	\begin{align}
		\by_{g}= &  \left(\bH_{\mathrm{d},g}^{\H}+\bH_{2,g}^{\H}\bPhi\bH_{1} \right)\bx+
		\bw_{g},\label{rec1}
	\end{align}
	where $\bH_{\mathrm{d},g}=[\bh_{\mathrm{d},g_{1}},\ldots,\bh_{\mathrm{d},g_{K_{g}}} ] \in \mathbb{C}^{M \times K_{g}}$  is the channel matrix describing the direct channels between the BS and the UEs in group $ g$ with $  \bh_{\mathrm{d},g_{i}}\in \mathbb{C}^{M \times 1}, $  $ i=1,\ldots,K_{g} $ being the individual UEs channels. Note that the index $ g_{i} $ denotes UE $ i $ in group $ g $. The subscripts $ 1$ and $ 2$ correspond to the BS-IRS and IRS-UEs  links, respectively. Hence, $ \bH_{2,g}= [\bh_{2,g_{1}},\ldots,\bh_{2,g_{K_{g}}} ]\in \mathbb{C}^{N \times K_{g}}$ describes the channel between the IRS and the $ g $th group with $  \bh_{2,g_{i}}\in \mathbb{C}^{N \times 1}, $  $ i=1,\ldots,K_{g} $. Also, $ \bH_{1}=[\bh_{1,1},\ldots,\bh_{1,N} ] \in \mathbb{C}^{M \times N}$ describes the LoS channel between the BS and the IRS with  $  \bh_{1,i}\in \mathbb{C}^{M \times 1}, $  $ i=1,\ldots,N $.     Moreover, $ \bx=\sum_{g=1}^{G} \bV_{g}\bd_{g}$, satisfying the power constraint $ \EE[\|\bx\|^{2}]=\tr\left(\bP\bV^{\H}\bV\right)\le P_{\mathrm{max}} $,  is the $ M\times 1 $ linearly precoded transmit signal vector, which  means that   $ \rho=P_{\mathrm{max}}/\sigma^{2} $ expresses the transmit SNR. Also,  $ \bd_{g}=\bP_{g}^{1/2} \bs_{g}\in \mathbb{C}^{  K_{g} \times 1} $, where  $ \bs_{g}\sim\mathcal{CN}\left(\b0,\Id_{K_{g}}\right)\in \CC^{K_{g} \times 1} $ expresses the vector of data symbols in group $ g $. Note that $ \bV=[\bV_{1}, \ldots,\bV_{G}] \in \mathbf{C}^{M\times K}$ and $\bP= \diag\left(\bP_{1},\ldots,\bP_{G}\right) $ with $ \bP_{g}=\diag\left(p_{g_{1}},\ldots,p_{g_{K_{g}}}\right)$ for $g=1,\ldots,G $, where  $ \bV_{g}\in \mathbb{C}^{M \times K_{g}} $  is the  linear precoding matrix for group $ g $ and $ P_{g}=\sum_{i=1}^{K_{g}}p_{g_{i}}$   is the total available transmit power for group $ g $ with  $ p_{g_{k}}\ge 0 $ expressing the signal power of UE $ k $. The vector $ \bw_{g} \sim \cC\cN\left(\b0,\sigma^2\Id_{K_{g}}\right) $ expresses the additive white Gaussian noise (AWGN) at the BS. Furthermore, $\bPhi=\mathrm{diag}\left( \al_{1}e^{j \phi_{1}}, \ldots, \al_{N}e^{j \phi_{N}} \right)\in\mathbb{C}^{N\times N}$ is the diagonal RBM, which  expresses the response of the $ N $ elements with $ \phi_{n} \in [0,2\pi]$ and $ \al_{n}\in [0,1] $ describing the phase  and amplitude coefficient for element $ n $, respectively. Henceforth, we make the common assumption of maximum  reflection ($ \al_{n}=1~ \forall n$) based on recent advances in  lossless metasurfaces \cite{Badloe2017}.  
	
	Despite the majority of existing works, e.g., \cite{Wu2019,Huang2019}, which assumed independent Rayleigh model, we consider the practical effect of spatial correlation, which is unavoidable in realistic IRS-assisted systems  \cite{Bjoernson2020}.\footnote{ Contrary to most works on IRS that model correlation by using models for conventional antenna arrays, we adopt the recently presented correlation model by the authors in \cite{Bjoernson2020}, being more suitable for  IRS.}  Also, according to the JSDM method, despite that the individual effects are different across different UEs, the covariances matrices, describing the aggregate result of path-loss and correlation,  have been assumed equal across different UEs of the same group since they are clustered accordingly. Thus, $ \bh_{\mathrm{d},g_{k}} $ and $ \bh_{2,g_{k}} $, concerning UE $ k $ in group $ g $, are expressed as
	\begin{align}
	\bh_{\mathrm{d},g_{k}}&=\bR_{\mathrm{BS},g}^{1/2}\bz_{\mathrm{d},g_{k}},~
		\bh_{2,g_{k}}=\bR_{\mathrm{IRS},g}^{1/2}\bz_{2,g_{k}},
	\end{align}
	where  $ \bR_{\mathrm{BS},g}=\beta_{\mathrm{d},g_{k}}\bR_{\mathrm{BS},g_{k}} \in \mathbb{C}^{M \times M} $ and $ \bR_{\mathrm{IRS},g}=\beta_{2,g_{k}}\bR_{\mathrm{IRS},g_{k}} \in \mathbb{C}^{N \times N} $ represent the deterministic Hermitian-symmetric positive semi-definite aggregate covariance 
	matrices at the BS and the IRS respectively  corresponding to group $ g $ with $ \tr\left(\bR_{\mathrm{BS},g} \right)=M $ and $ \tr\left(\bR_{\mathrm{IRS},g} \right)=N $. Note that $ \beta_{\mathrm{d},g_{k}} $, $ \bR_{\mathrm{BS},g_{k}} $ and $ \beta_{2,g_{k}}$, $ \bR_{\mathrm{IRS},g_{k}}  $ are the path-losses, correlations of  the BS-UE $ k $  and IRS-UE $  k $ links in group $ g $, respectively. This channel modeling is very versatile since it can account for both  correlation ($\bR_{\mathrm{BS},g_{k}}  ,\bR_{\mathrm{IRS},g_{k}}$) and path loss ($ \beta_{\mathrm{d},g_{k}},\beta_{2,g_{k}} $) simultaneously or either of the two. Certain estimation methods (see e.g., \cite{Neumann2018,Upadhya2018}) allow to obtain the correlation matrices and the path-losses,
	which can, thus, be  assumed  known by the network. Moreover, $ \bz_{\mathrm{d},g_{k}} \sim \mathcal{CN}\left(\b0,\Id_{M}\right) $ and $ \bz_{2,g_{k}}\sim \mathcal{CN}\left(\b0,\Id_{N}\right) $ express the corresponding fast-fading vectors.

The  channel matrix $ \bH_{1} $ can be expressed as
	\begin{align}
		[\bH_{1}]_{m,n}&=\sqrt{\beta_{1}} \exp\Big(j \frac{2 \pi }{\lambda}\left(m-1\right)d_{\mathrm{BS}}\sin \theta_{1,n}\sin \phi_{1,n}\nn\\
		&+\left(n-1\right)d_{\mathrm{IRS}}\sin \theta_{2,m}\sin \phi_{2,m}\Big)\!.
	\end{align}
	Here, $ \lambda $ is the carrier wavelength, $ \beta_{1} $ is the path-loss between the BS and the IRS. Also,  $ d_{\mathrm{BS}} $ and $ d_{\mathrm{IRS}} $ are the inter-antenna separation at the BS and inter-element separation at the IRS, respectively \cite{Nadeem2020}. The elevation and azimuth LoS angles of departure (AoD) at the BS with respect to IRS element $ n $ are described by $ \theta_{1,n} $ and $ \varphi_{1,n} $. The elevation and azimuth LoS angles of arrival (AoA) at the IRS are described by $ \theta_{2,n} $ and $ \varphi_{2,n} $. 	The phases corresponding to the response of the surface elements  adjusted in a way so that  $ \rank(\bH_{d,g}^\H+\bH_{2,g}^\H\bPhi \bH_1)  \ge K $, which is the necessary condition for supporting multi-stream transmission to the K users of a group. This approach of rank improvement through an intelligent surface has been proposed and validated in \cite{Oezdogan2020} for a scenarios with a 2-antenna BS and a 2-antenna UE. In our case, we generalize this approach to a system with an arbitrary number of UEs.
	
	Given the RBM, the overall channel vector between the BS and UE $ k $ in  group $  g $ is written according to \eqref{rec1} as $ \bh_{g_{k}}=\bh_{\mathrm{d},g_{k}}+\bH_{1}\bPhi\bh_{2,g_{k}}$, which is distributed as $ \bh_{g_{k}}\sim \cC\cN\left( \b0, \bR_{g} \right) $ with $ \bR_{g}= \bR_{\mathrm{BS},g}+ \bH_{1} \bPhi {\bR}_{\mathrm{IRS},g}\bPhi^{\H}\bH_{1}^{\H}$.
	
	It is worthwhile to mention that according to the FDD design with training for channel estimation in both the uplink and the downlink, the following procedure takes place, which consists of two frames. 	In the uplink frame, each UE first transmits orthogonal 	pilot symbols to the BS, where these are used to 	estimate the underlying channels. Following this, all the UEs 	simultaneously transmit their data to the BS. The data 	is detected at the BS using the CSI acquired during the pilot transmission phase.  During the downlink training 	frame, the BS transmits orthogonal pilot symbols 	from its antennas which are used by the UEs to estimate 	the corresponding downlink channel. These channel estimates 	are then fed back to the BS by the UEs, and are used by the latter to appropriately precode the data symbols 	transmitted simultaneously to all the UEs during the data	transmission phase.

In terms of implementation, two approaches could be used that present a trade-off between performance and cost. The first approach assumes that makes no compromise on the performance but has a higher cost. It requires $ 2 N$ IRS elements, where $ N $ elements are used for the uplink transmission and $ N $ elements are used for the downlink transmission. The downlink/uplink group of elements operates at the corresponding uplink/downlink frequency. Note that this approach requires all the corresponding hardware to be doubled.\footnote{It  is true that the gap between the downlink and uplink in 4G/5G systems is not big. It is even lower than 0.3 GHz. For instance, in LTE Band 1 the uplink is defined in the range $ 1.92-1.98 $ GHz, whereas the downlink is defined in the range $ 2.11-2.17 $ GHz \cite{Radio2011}”. Hence the gap between the uplink and downlink is only$  0.13 $ GHz. However, please have in mind that normally the unit cells of the metasurface that a RIS consists of elements having a narrowband frequency response, unless special designs are manufactured that create a wideband response. For instance, in Section II  of \cite{You2022} it is stated that: “For instance, reflective metasurfaces composed of microstrip patch resonators typically exhibit a fractional bandwidth of less than $ 5 $\%.” Even if we assume the worst-case scenario of $ 5 $\% fractional bandwidth, this would mean that the frequency response in a downlink FR2 LTE case with carrier frequency $ 2.14 $ GHz is limited in the range $  0.05*2.14$ GHz$=0.107 $ GHz, which is smaller than the aforementioned 0.13 GHz gap between the uplink and downlink in LTE. Hence, with the proper design of the intelligent surface unit cells, the frequency response of the unit cells dedicated for the two bands, uplink and downlink, can be such that there is no interference from one band to the other. }\footnote{The proposed design concerns 5G and beyond networks, where possibly the frequency interval between the uplink and  downlink will be greater than $ 0.3 $ GHz, which is in 4G/5G systems. Hence, there will be no coupling or interference between the uplink and downlink. Despite this, even for smaller intervals, there is no issue because of the following reasons. Doubling the hardware means that there will be separated beamforming networks/control circuits for the uplink RIS and downlink RIS; thus, the uplink RIS and downlink RIS can operate independently. We would like to mention that similar techniques have been well-developed for the phased array design. For example, in \cite{Lee2018}, an architecture of using a separate transmitter and receiver  array operating at $ 94 $ GHz is presented.  Second, RF filters with high selectivity are used to suppress any unwanted signals from other channels belonging to other frequencies. } Without significant performance loss, 			the whole procedure can be  facilitated by the use of diplexers that allow the simultaneous processing of two different frequencies in common FDD systems and also enable the separate phase shifts optimizations that happen during the uplink and downlink transmissions.  In particular, the use of radio-frequency (RF)-micro-electromechanical system (MEMS) single-pole double-throw (SPDT) switches to realize phased array transceivers has been well-developed in antenna engineering with applications from microwave frequency to mm-wave frequency ranges. As shown Fig. 2 in \cite{Chaloun2014}, the uplink/downlink  links, can be realized on a single monolithic microwave integrated circuit (MMIC) with a shared vector modulator (where the phase shifter is located). This well-developed technology is compatible with IRS design and can be applied to IRS to support simultaneous FDD channels. Notably,  the switching time of modern SPDT is in the scale of picoseconds. Thus, the uplink and downlink signals can be recovered as far as the speed of the switch can provide enough samples to recover the signals, which is possible according to \cite{Chaloun2014}. Moreover, another technique is  through the use of a microwave circulator \cite{Dinc2017}, where the transmit signal will be delivered to the antenna while the received signal from the same antenna will only be delivered to the receiver port.

	\subsection{Advantages/disadvantages of FDD in IRS-assisted systems}
	FDD has several advantages over TDD that become more pronounced in IRS-assisted systems. For example,  FDD design is accompanied with lower latency since  communication takes place simultaneously while in TDD there is a switching between transmission and reception. Also, in IRS-assisted systems, where the distance between the transmitter and the receiver is increased, the guard period in TDD systems, being proportional to the distance, increases and affects the performance, while FDD does not appear a problem with large distances. Furthermore, the medium access control (MAC) layer of TDD systems is more complex because it requires accurate time synchronization between uplink and downlink, while FDD systems do not require any uplink/downlink switching mechanism at timescale. However, FDD systems come with higher implementation cost demands such as the diplexers that are required to make the operation in different frequencies possible. Moreover, given that in practical systems, most network volume is consumed in the downlink, in TDD, it is possible to balance the traffic between uplink and downlink  by utilising more time slots for downlink. In addition, FDD, using two frequencies instead of one used in TDD, allocates more spectrum. 
	
	\subsection{PGP approach}
	The JSDM approach was presented in terms of two methods, the joint-group processing (JGP) and PGP \cite{Adhikary2013}. JGP can be applied in the case that the estimation and feedback of the whole matrix from all UEs in all groups are affordable. Normally, the JGP method is accompanied by prohibitive overhead for channel estimation and feedback while PGP is more advantageous due to lower overhead. For this reason, the PGP method, summarized below, is suggested as a more practical approach and is employed in this work.

	\subsubsection{Signal model}	 PGP suggests a two-stage precoding matrix for   group $ g $, resulting in feedback overhead and the computational complexity reductions,  expressed as 
	\begin{align}
		\bV_{g}=\bB_{g}\bF_{g}, \label{prec1}
	\end{align}
	where $ \bB_{g}\in \CC^{M \times b_{g}} $ is the preprocessing matrix based on the long-term channel statistics with $ b_{g}\ge K_{g} $ being  an integer parameter to be optimized.   Also, $ \bF_{g}\in \CC^{b_{g} \times K_{g}} $, $ g=1,\ldots,G $ is the  precoding matrix based on the instantaneous channel (short-term CSI) of group $ g $.
	
	
	
	Hence, the received signal in \eqref{rec1} with dual precoding described by  \eqref{prec1} can be written  as
	\begin{align}
		\by_{g}= &  \bH_{g}^{\H}\bB_{g}\bF_{g}\bd_{g}+ \sum_{i \ne g}^{G}\bH_{g}^{\H}\bB_{i}\bF_{i}\bd_{i}+
		\bw_{g},\label{rec2}
	\end{align}
	where $\bH_{g}=[\bh_{g_{1}},\ldots,\bh_{g_{K_{g}}} ] \in \mathbb{C}^{M \times K_{g}}$. 
	
	Based on the Karhunen-Loeve representation, the channel vector $ \bh_{g_{k}} $ can be written as
	\begin{align}
		\bh_{g_{k}}=\bU_{g}\bLam_{g}^{\frac{1}{2}}\bz_{g_{k}},\label{Kar}
	\end{align}
	where $ \bU_{g}\in \CC^{M\times r_{g}} $ is a tall unitary matrix including the eigenvectors of $ \bR_{g} $ that correspond to the non-zero eigenvalues,  $\bLam_{g}  $ is an $ r_{g} \times r_{g} $ diagonal matrix with elements the non-zero eigenvalues of $\bR_{g}  $, and  $ \bz_{g_{k}} \in \CC^{r_{g} \times 1} \sim \mathcal{CN}\left(\b0,\Id_{r_{g}}\right)  $. Generally, $ M\gg r_{g} $. Thus, we have $ \bR_{g}=\bU_{g}\bLam_{g}\bU_{g}^{\H} $.

	\subsubsection{Imperfect CSI}	Certain reasons such as quantization can result in imperfect feedback of the CSI to the BS (see Rem. \ref{rem1} below), which is accompanied with significant overhead. Especially, the imperfect CSI at the BS from group $ g $ can be expressed as
	\begin{align}
		\hat{\bZ}_{g}=\sqrt{1-\tau_{g}^2}\bZ_{g}+\tau_{g}\bE_{g},\label{imp}
	\end{align}
	where $ \bZ_{g}=[\bz_{g_{1}, \ldots},\bz_{g_{K_{g}}}]\in \mathbb{C}^{M \times K_{g}}$ is the perfect CSI and $\bE_{g} =[\bee_{g_{1}, \ldots},\bee_{g_{K_{g}}}]\in \mathbb{C}^{M \times K_{g}} $ with $ \bee_{g_{k}}  $ denoting the corresponding error that consists of elements with zero mean and unit variance. The parameter $ \tau_{g}\in [0,1] $ describes the accuracy of the available CSI in group $ g $. Hence, if $ \tau_{g}=0 $, we obtain perfect CSI while $ \tau_{g}=1 $ implies no CSI knowledge. Without loss of generality, we assume that the levels of CSI accuracy for all groups are the same, i.e., $ \tau_{1}=\ldots=\tau_{G}=\tau $. From \eqref{imp}, $	\hat{\bH}_{g}$ can be defined as the imperfect CSI knowledge of 	$ \bH_{g} $ by using 	$ \hat{\bZ}_{g} $. Note that the imperfect CSI $\hat{\bH}_{g}$, given in terms of $ {\bH}_{g} $ is based on a given RBM.  The assumption of a fixed RBM during  one coherence interval is justified based on the fact that RBM is expected to change according to large-scale statistics expressed by $ \bR_{g} $, which changes at every several coherence intervals.
	
	\begin{remark}\label{rem1}
		Having assumed that both the BS and UEs are aware of large-scale statistics, the model in \eqref{imp} can describe the scenario, where  the $ k $th UE in the group $ g $ quantizes $ \bh_{g_{k}} $ with the help of a random codebook and sends back the codeword index  to the BS \cite{Jindal2006}. We provide an example of the advantage of using this method in Sec. \ref{Numerical}.
	\end{remark}

		\subsubsection{Design of pre-beamforming} Using block diagonalization (BD) preprocessing  to cancel out the inter-group interference, we design the pre-beamforming matrix as
	\begin{align}
		\bH_{g}^{\H}\bB_{i}\approx 0,~~\mathrm{for}~i\ne g,
	\end{align}
	where the index $ i $ corresponds to  group $i $. In the case of exact BD, \eqref{rec2} becomes
	\begin{align}
		\by_{g}= &  \bH_{g}^{\H}\bB_{g}\bF_{g}\bd_{g}+ 
		\bw_{g}.\label{PGP1}
	\end{align}

	Exact BD can be achieved, if $\mathrm{Span}(\bU_{g}) \nsubseteq \mathrm{Span}(\{\bU_{i}:i\ne g\})$ while the necessary multiplexing gain can be achieved if $ \dim\left(\mathrm{Span}\left(\bU_{g}\right)\cap \mathrm{Span^{	\bot}\left(\{\bU_{i}: i\ne g\}\right)}\right)\ge K_{g} $ for all groups $ g =1, \dotsb, G$. In other words, for given sets of $ \{b_{g}\} $ and $ \{K_{g}\} $ satisfying $ K_{g}\le b_{g}\le r_{g} $, $ \bB_{g} $ can be designed if $ \bU_{i}^{\H} \bB_{g}=0$ for all $ i\ne g $ and $ \rank(\bU_{g}^{\H}\bB_{g})\ge K_{g} $. If the dimension of $ \mathrm{Span}\bot(\{\bU_{i}:i\ne g\}) $ is less than $ K_{g} $, the rank condition is not fulfilled. In such case, there are two options: i) reduction of $ K_{g} $, or ii) resort to approximate BD.
	
Herein,  we choose approximate BD to design $ \{\bB_{g}\} $. For this reason, we choose  $ r_{g}^{\star}\le r_{g}  $   dominant eigenvalues of  $\bR_{g}  $. On this ground, we write $ \bU_{g}=[ \bU_{g}^{\star}, \bU_{g}^{'}] $, where $ \bU_{g}^{\star} $ is the $ M \times  r_{g}^{\star} $ matrix consisted of the dominant eigenvectors and $  \bU_{g}^{'} $ is the  $M \times  (r_{g}-r_{g}^{\star}) $ matrix including the eigenvectors corresponding to the weakest eigenvalues. To achieve  approximate BD, we have to choose the dominant eigenmodes $ r_{g}^{\star} $ for each group $ g $   subject to the condition $\mathrm{Span}(\bU_{g}^{\star}) \nsubseteq \mathrm{Span}(\{\bU_{i}^{\star}:i\ne g\})$ for all groups $ g =1, \dotsb, G$. Notably, the design of $ \{\bB_{g}\} $ has to take into   account  a feasible choice of the parameters $ \{r_{g}^{\star}\} $, $ \{b_{g}\} $, and $ \{K_{g}\} $ that should be optimized to maximize the SE under a given system setup. 	The matrix $ \bB_{g} $ is constructed as in \cite{Adhikary2013}. Herein, we briefly summarize the main steps.  Hence, first, we define the matrix
	\begin{align}
		\bU_{-g}=[\bU_{1}^{\star},\ldots,\bU_{g-1}^{\star},\bU_{g+1}^{\star}, \ldots, \bU_{G}^{\star}],
	\end{align}
	whose size is $ M \times \sum_{i\ne g} r_{i}^{\star}$ and its rank is $\sum_{i\ne g} r_{i}^{\star}  $. Based on SVD, let a set of left eigenvectors of $ \bU_{-g} $ denoted by $ [\bE_{g}^{(1)},\bE_{g}^{(0)}] $, where $ \bE_{g}^{(1)} $ expresses the left singular vectors corresponding to the $ \sum_{i\ne g} r_{i}^{\star} $ dominant singular values $\bLam_{g}^{(1)}  $ while $ \bE_{g}^{(0)}  $  corresponds to  the  left singular vectors with $M- \sum_{i\ne g} r_{i}^{\star} $ non-dominant singular values $\bLam_{g}^{(0)}  $. Now, we define the matrix $ \tilde{\bH}_{g}=(\bE_{g}^{(0)})^{\H}\bH_{g} $, which has the property to be orthogonal  to the dominant eigenspace spanned by the channel of  other groups. The covariance matrix of $ \tilde{\bH}_{g} $ is given by
	\begin{align}
		\tilde{\bR}_{g}&=(\bE_{g}^{(0)})^{\H}\bR_{g}\bE_{g}^{(0)}	\label{kar3}\\
		&=(\bE_{g}^{(0)})^{\H}\bU_{g}\bLam_{g}\bU_{g}^{\H}\bE_{g}^{(0)}\label{kar1}\\
		&=\bG_{g}\bQ_{g}\bG_{g}^{\H},	\label{kar2}
	\end{align}
	where \eqref{kar1} is based on \eqref{Kar} while \eqref{kar2} is the SVD of $ \tilde{\bR}_{g} $, i.e., $ \bG_{g} $ describes the eigenvectors of $ \tilde{\bR}_{g} $. The pre-beamforming matrix $ \bB_{g} $ is obtained by setting $ \bG_{g}=[\bG_{g}^{(1)},\bG_{g}^{(0)}] $, where $ \bG_{g}^{(1)} $ corresponds to  the dominant $ b_{g} $ eigenmodes of 	$ \tilde{\bR}_{g} $. In particular, we have that
	\begin{align}
		\bB_{g}=\bE_{g}^{(0)}\bG_{g}^{(1)},
	\end{align}
	which means that $ \bB_{g} $ coincides  with the $ b_{g} $ dominant eigenmodes of $ \tilde{\bR}_{g} $ while it is orthogonal to the dominant $ r_{i}^{\star} $ eigenmodes of groups $ i\ne g $. Gathering all constraints, the  condition  $ K_{g}\le b_{g} \le \rank(\tilde{\bR}_{g})$, where $ \rank(\tilde{\bR}_{g})=\min\{M- \sum_{i\ne g} r_{i}^{\star} ,r_{g}\} $ should be satisfied.  

	\section{Asymptotic Performance Analysis with Dual Precoding}\label{AsymptoticPerformance}
	In this section, we first provide the analysis towards the derivation of the downlink achievable SE  of IRS-assisted systems based on FDD while  employing the JSDM method in the case of the PGP approach. 
	
	The derivation of the DE SINR requires the following assumptions concerning the correlation matrices and the power allocation matrix as in \cite{Wagner2012}.
	
	\begin{assumption}\label{As1}
		All correlation matrices, i.e.,  $ \bR_{\mathrm{IRS},g} $ and $ \bR_{\mathrm{BS},g}, $ $ g=1,\ldots,G $  have uniformly	bounded spectral norm on $ N $ and $ M $, respectively, i.e.,
		\begin{align}
			&	\lim_{N,K\to \infty}\!\!\!\! \sup\sup_{1\le k\le K}\|\bR_{\mathrm{IRS},g} \|<\infty,~~~~
				\lim_{M,K\to \infty}\!\!\!\! \sup\sup_{1\le k\le K}\|\bR_{\mathrm{BS},g} \|<\infty
		\end{align}
	\end{assumption}

	\begin{assumption}\label{As2}
		The maximum  power of group $  g $, i.e.,  $ \max\left(p_{g1},\ldots, p_{gK_{g}}\right) $ is of the order $ \mathcal{O}\left(1/K_{g}\right) $, being equivalent to
		\begin{align}
			\|\bP_{g}\|=\mathcal{O}\left(1/K_{g}\right).
		\end{align}
	\end{assumption}

	According to \eqref{rec2}, the effective channel of group $ g $ is $ \bar{\bH}_{g}=\bB_{g}^{\H}\bH_{g} $ with covariance matrix for UE $ g_{k} $ being $ \bar{\bR}_{g}= \bB_{g}^{\H}\bR_{g}\bB_{g}$. For the sake of exposition, we assume  the same number of dominant eigenvalues per group, the same number of UEs per group, and the same dimension of  the pre-beamforming matrix per group, i.e., $ r_{1}^{\star}=\ldots=r_{G}^{\star}=\bar{r} $, $K_{1}=\ldots=K_{g}= {\bar{K}} $, and $ b_{1}=\ldots=b_{g}={\bar{b}} $. The extension to the general case is immediate and straightforward. Hence, the dimensions of $ \bar{ \bH}_{g} $ are $  {\bar{b}}\times  {\bar{K}} $.  The precoding matrix for group $ g $  is designed in terms of the instantaneous imperfect CSI to cancel out the intra-group interference on that group as
	\begin{align}
		\bF_{g}=\sqrt{\lambda_{g}}{\bSigma}_{g}\hat{\bar{\bH}}_{g},
	\end{align}
	where $ {\bSigma}_{g}=\left(\hat{\bar{\bH}}_{g}\hat{\bar{\bH}}^{\H}_{g}+\bar{b} \al  \Id_{\bar{b}}\right)^{-1} $ with   $\hat{\bar{\bH}}_{g}=\bB_{g}^{\H}\hat{\bH}_{g} $ being the effective channel estimate at the BS while $ \hat{\bH}_{g} $ is the  imperfect CSI knowledge of $ \bH_{g} $. Also  $ \al  $, selected as $ \al=\frac{M}{\bar{b}P_{\mathrm{max}}} $ to be equivalent with the RZF
	linear filter \cite{Tse2005},  is a regularization factor that makes expressions converge to a constant, and $ \lambda_{g} $ is the normalization parameter that fulfils the power constraint for group $ g $.  In particular, the normalization parameter becomes
	\begin{align}
		\lambda_{g}&=\frac{P_{\mathrm{max}}}{\tr(\bP\hat{{\bH}}_{g}^{\H}\bSigma_{g}\bB_{g}^{\H}\bB_{g}\bSigma_{g}\hat{{\bH}}_{g})}\nn\\
		&=\frac{P_{\mathrm{max}}}{\tr(\bP_{g}\hat{{\bH}}_{g}^{\H}\bSigma_{g}^{2}\hat{{\bH}}_{g})}\label{lam2}\\
		&=\frac{P_{\mathrm{max}}}{\Psi_{g}},\label{lam21}
	\end{align}
	where \eqref{lam2} is obtained because $ \bB_{g}^{\H}\bB_{g}=\Id_{\bar{b}} $ as the product of two tall matrices. Hence, based on \eqref{rec2} and \eqref{lam21}, the SINR of the $ k $th UE in group $ g $ with perfect receiver CSI 	is given by
	\begin{align}
		\gamma_{g_{k},\mathrm{PGP}}=\frac{	\mathrm{DS}_{g_{k}}}{ \mathrm{SGI}_{g_{k}}+\mathrm{IGI}_{g_{k}}+\sigma^{2}},
	\end{align}
	where 
	\begin{align}
		\mathrm{DS}_{g_{k}}&=p_{g_{k}}\lambda_{g}|\bh_{g_{k}}^{\H}\bB_{g} \bSigma_{g}\hat{\bar{\bh}}_{g_{k}}|^{2},\\
		\mathrm{SGI}_{g_{k}}&=\lambda_{g}\sum_{j\ne k}p_{g_{j}}|\bh_{g_{k}}^{\H}\bB_{g} \bSigma_{g}\hat{\bar{\bh}}_{g_{j}}|^{2},\\
		\mathrm{IGI}_{g_{k}}&=\sum_{l\ne g}\sum_{j}\lambda_{l}p_{lj}|\bh_{g_{k}}^{\H}\bB_{l} \bSigma_{l}\hat{\bar{\bh}}_{lj}|^{2}
	\end{align}
	with the numerator expressing the desired signal power while the first  and second terms in  the denominator express the self-group and inter-group interferences,  respectively. 
	
	Under  the Assumptions \ref{As1}-\ref{As2} and by assuming that $ M \to \infty $ while  $ \bar{K} $, $ \bar{r} $, and $ \bar{b} $  also go to infinity by keeping their ratio with $ M $ fixed, the DE SINR $ \bar{\gamma}_{g_{k},\mathrm{PGP}} $ fulfils
	\begin{align}
		\gamma_{g_{k},\mathrm{PGP}}-\bar{\gamma}_{g_{k},\mathrm{PGP}}\xrightarrow[M \rightarrow \infty]{\mbox{a.s.}}0.\label{DE10}
	\end{align}
	
	\begin{theorem}\label{theorem:PGP}
		The  DE of the downlink SINR of UE $g_{k}$ with PGP in IRS-assisted mMIMO  systems with FDD, accounting for imperfect CSI, is given by
		\begin{align}
			\bar{\gamma}_{g_{k},\mathrm{PGP}}=\frac{\bar{S}_{g_{k}}}{\bar{I}_{g_{k}}}
		\end{align}
		where $\bar{S}_{g_{k}}= p_{g_{k}}\left(1-\tau^{2}\right)\bar{\delta}_{g}^{2} $ and $\bar{I}_{g_{k}}= \bar{\bY}_{gg}\left(1-\tau^{2}\left(1-\left(1+\bar{\delta}_{g}^{2}\right)^{2}\right)\right)+\left(1+\sum_{l\ne g}\bar{\lambda}_{l}\bar{\bY}_{gl}\right)\frac{\left(1+\bar{\delta}_{g}\right)^{2}}{\bar{\lambda}_{g}} $ with $ \bar{\lambda}_{g}=\frac{\rho}{\bar{\Psi}_{g}} $. 
		The expressions of  $\bar{ \delta}_{g} $, $ \bar{\Psi}_{g} $, $ \bar{\bY}_{gg} $, $ \bar{\bY}_{gl} $ are the unique solutions of\\ 
		\begin{align}
			\bar{\delta}_{g}&=\frac{1}{\bar{b}}\tr\left(\bar{\bR}_{g}\bT_{g}\right),~\bT_{g}=\left(\frac{\bar{K}}{\bar{b}}\frac{\bar{\bR}_{g}}{1+\bar{\delta}_{g}}+\al \Id_{b}\right)^{-1},\label{s1}
			\\
			~\bar{\Psi}_{g}&=\frac{P_{g}}{\bar{b}}\frac{\bar{m}_{g}}{\left(1+\bar{\delta}_{g}\right)^{\!2}},  \bar{\bY}_{gg}\!=\! \frac{P_{g}}{\bar{b}}\!\!\left(\!\!1-\frac{p_{g_{k}}}{P_{g}}\!\!\right)	\frac{\bar{m}_{gg}}{\left(1+\bar{\delta}_{g}\right)^{2}},\\
			  \bar{\bY}_{gl} &= \frac{1}{\bar{b}}	\sum_{k=1}^{\bar{K}}p_{l_{k}}\frac{m_{gl}}{\left(1+\bar{\delta}_{l}\right)^{2}},~\bar{m}_{g}=\frac{\frac{1}{\bar{b}}\tr\left(\bar{\bR}_{g}\bT_{g}\bB_{g}^{\H}\bB_{g}\bT_{g}\right)}{1-\frac{\frac{\bar{K}}{\bar{b}}\tr\left(\bar{\bR}_{g}\bT_{g}\bar{\bR}_{g}\bT_{g}\right)}{\bar{b}\left(1+\bar{\delta}_{g}\right)^{2}}},\\ \bar{m}_{gg}&\!=\!\frac{\frac{1}{\bar{b}}\!\tr\!\left(\bar{\bR}_{g}\bT_{g}\bar{\bR}_{g}\bT_{g}\right)}{1\!-\!\frac{\frac{\bar{K}}{\bar{b}}\tr\left(\bar{\bR}_{g}\bT_{g}\bar{\bR}_{g}\bT_{g}\right)}{\bar{b} \left(1+\bar{\delta}_{g}\right)^{2}}},~\!
			\bar{m}_{gl}\!=\!\frac{\frac{1}{\bar{b}}\!\tr\!\left(\bar{\bR}_{l}\bT_{l}\bB_{l}^{\H}{\bR}_{g}\bB_{l}\bT_{l}\right)}{1-\frac{\frac{\bar{K}}{\bar{b}}\tr\left(\bar{\bR}_{l}\bT_{l}\bar{\bR}_{l}\bT_{l}\right)}{\bar{b}\left(1+\bar{\delta}_{l}\right)^{2}}}.\label{s6}
		\end{align}
	\end{theorem}
	\proof The proof  is provided in Appendix~\ref{theorem1}.\endproof
	
	Notably,  Assumption \ref{As2} allows the omission of term $ \frac{p_{g_{k}}}{P} $ in $ \bar{\bY}_{gg} $ since the  convergence in \eqref{DE10} is still true.

	Based  on the dominated convergence~\cite{Billingsley2008} and the continuous mapping theorem~\cite{Vaart2000}, the DE of the sum-rate  $ \sum_{g=1}^{G}\sum_{k=1}^{K}\log_{2}(1 + 	{\gamma}_{g_{k},\mathrm{PGP}}) $ is given by
	\begin{align}
		\mathrm{\overline{SR}}_{\mathrm{PGP}}=\sum_{g=1}^{G}\sum_{k=1}^{K}\log_{2}(1 + 	\bar{\gamma}_{g_{k},\mathrm{PGP}}), \label{SUM_PGP}
	\end{align}
	where $ 	\sum_{g=1}^{G}\sum_{k=1}^{K}\log_{2}(1 + 	{\gamma}_{g_{k},\mathrm{PGP}})-\mathrm{\overline{SR}}_{\mathrm{PGP}}\xrightarrow[ M \rightarrow \infty]{\mbox{a.s.}}0. $
	
	The nature of the multicast mechanism indicates that the  DE sum-rate of group $ g $, allocated with total power $ P_{g} $ such that $ \sum_{g=1}^{G}P_{g}=P_{\mathrm{max}} $, is determined by $ \mathrm{\overline{SR}}_{\mathrm{PGP},g}=\sum_{k=1}^{K}\mathrm{\overline{SR}}_{\mathrm{PGP},g_{k}}$, where $ \mathrm{\overline{SR}}_{\mathrm{PGP},g_{k}}=\log_{2}(1 + 	\bar{\gamma}_{g_{k},\mathrm{PGP}})  $ is the DE rate of UE $ k $ in group $ g $.
	\begin{proposition}\label{Prop:groupPower}
		In IRS-assisted mMIMO systems with PGP, all UEs in group $ g $  are allocated with equal power $ P_{g}/\bar{K} $ and the DE sum-rate of group $ g $ is 	$ \mathrm{\overline{SR}}_{\mathrm{PGP},g}=\bar{K} R^{\star}_{g}$ since all these UEs have equal achievable rate $ R^{\star}_{g} $.
	\end{proposition}
	\proof The proof  is provided in Appendix~\ref{Prop1}.\endproof
	\begin{remark}
		The results, presented by Prop. \ref{Prop:groupPower}, rely on the fact that all UEs in the group $ g $ have the same covariance matrix.
	\end{remark}

	\begin{corollary}\label{cor1}
		Based on	Proposition \ref{Prop:groupPower}, the DE rate of UE $ k $ in group $ g $ can be written as
		\begin{align}
			\bar{\gamma}_{g,\mathrm{PGP}}=\frac{\bar{S}_{g}}{\bar{I}_{g}},
		\end{align}
		where $\bar{S}_{g}=\frac{P_{g}}{\bar{K}}\left(1-\tau^{2}\right)\bar{\delta}_{g}^{2} $ and $\bar{I}_{g}= \bar{\bY}_{gg}\big(1-\tau^{2}\big(1-\left(1+\bar{\delta}_{g}^{2}\right)^{2}\big)\big)+\big(1+\sum_{l\ne g}\bar{\lambda}_{l}\bar{\bY}_{gl}\big)\frac{\left(1+\bar{\delta}_{g}\right)^{2}}{\bar{\lambda}_{g}} $ with 
		$ 
		\bar{\bY}_{gg} = \frac{P_{g}}{\bar{b}}\left(1-\frac{1}{\bar{K}}\right)	\frac{\bar{m}_{gg}}{\left(1+\bar{\delta}_{g}\right)^{2}} $ and  	 	
		$ \bar{\bY}_{gl} = \frac{1}{\bar{b}}	\frac{P_{l}}{\bar{K}}\frac{m_{gl}}{\left(1+\bar{\delta}_{l}\right)^{2}} $  while all other variables, given by Theorem \ref{theorem:PGP}, remain the same.
	\end{corollary}
	\section{Sum SE Maximization Design}\label{SumSEMaximizationDesign}
	In this section, we perform optimization of sum SE of the overall IRS-assisted mMIMO  system. In particular, the sum-rate maximization problem is expressed as follows
	\begin{subequations}
		\begin{align}
			(\mathcal{P}2)~~&\max_{\bPhi,\bp\ge 0} 	\;		\mathrm{\overline{SR}}_{\mathrm{PGP}}=\max_{\bPhi,\bp}\bar{K}\sum_{g=1}^{G}\log_{2}(1 + 	\bar{\gamma}_{g,\mathrm{PGP}})
			\label{Maximization1} \\
			&~	\mathrm{s.t}~~~\;\!\sum_{g=1}^{G}P_{g}\le P_{\mathrm{max}},	\label{Maximization2} \\
			&\;\quad\;\;\;\;\;\!\!~\!|\tilde{\phi}_{n}|=1,~~ n=1,\dots,N,
			\label{Maximization3} 
		\end{align}
	\end{subequations}
	where constraint \eqref{Maximization2} guarantees that the BS transmit power is kept below the maximum power $ P_{\mathrm{max}} $ and constraint \eqref{Maximization3} expresses that each IRS element results in only a phase shift without any amplification of the incoming signal. Note that for the sake of exposition, we have defined the vector $ \bp=[P_{1}, \ldots,P_{G}]^{\T} $ and  $\tilde{\phi}_{n}= \exp\left(j \phi_{n}\right) $ for all $ n $ corresponding to the elements of $ \bPhi $.
	\begin{corollary}
		Let $ \bR_{\mathrm{IRS},g}= \Id_{N}$, then $ \bR_{g} $ does not depend on $ \bPhi $. Equivalently,  $ \bar{\bR}_{g} $ is independent on the phase shifts, and thus, $ \bar{\gamma}_{g,\mathrm{PGP}}  $ cannot be optimized in this case.
	\end{corollary}

	The optimization problem $ 	(\mathcal{P}2) $ is non-convex and subject to  a unit-modulus constraint regarding $ \tilde{\phi}_{n} $, which make its solution challenging. To address this, we rely on the common approach in IRS literature, being the application of the alternating optimization (AO) technique, where $ \bPhi $
	and  $ \bp $ are going to be solved separately and iteratively. Hence, first, we solve for $ \bPhi $ given a fixed $ \bp $.  Next, we focus on finding  the optimum $ \bp $ with $ \bPhi $ fixed. The iteration of this process achieves the increase of $ \mathrm{\overline{SR}}_{\mathrm{PGP}} $ at each iteration step until convergence of the objective to  its optimum value  since it is upper-bounded  due to the power constraint \eqref{Maximization2}.  


	\subsection{IRS Design}
	So far, we have considered the RBM fixed. However, to exploit the IRS towards the maximization of the sum SE, the RBM has to be optimized. Notably, we observe its presence inside the covariance matrices that appear in the DE expression of the SE with PGP. By assuming infinite resolution phase shifters and under imperfect CSI conditions, the RBM optimization problem for single-cell FDD systems employing the  JSDM method is formulated as
	\begin{align}\begin{split}
			(\mathcal{P}3)~~~~~~~\max_{\bPhi} ~~~	&		\mathrm{\overline{SR}}_{\mathrm{PGP}}\\
			\mathrm{s.t}~~~&|\tilde{\phi}_{n}|=1,~~ n=1,\dots,N,
		\end{split}\label{Maximization} 
	\end{align}
	where  
	$\mathrm{\overline{SR}}_{\mathrm{PGP}} $ is given by  \eqref{Maximization1}.  The maximization problem  $ 	(\mathcal{P}3) $ is  non-convex with respect to $ \bPhi $ while it is subject to  a unit-modulus constraint regarding $ \tilde{\phi}_{n} $. A local optimal solution to this problem can be obtained by applying the projected gradient 	ascent algorithm until converging to a 	stationary point, which means that, at every step, we project the solution onto the closest feasible point satisfying the unit-modulus constraint concerning $ \tilde{\phi}_{n} $ \cite{Kammoun2020}. Specifically, at step $ l $, the phases are included in the vectors $ \bs_{l} =[\tilde{\phi}_{l,1}, \ldots, \tilde{\phi}_{l,N}]^{\T}$. 
	The next  step of the iteration towards  convergence increases  $\mathrm{\overline{SR}}_{g_{k},\mathrm{l}} $, and is described by
	\begin{align}
		\tilde{\bs}_{l+1}&=\bs_{l}+\mu \bq_{l},~
		\bs_{l+1}=\exp\left(j \arg \left(\tilde{\bs}_{l+1}\right)\right),\label{sol2}
	\end{align}
	where the parameter $ \mu $ describes the step size and $ \bq_{l} $ expresses the  ascent direction at step $ l $.  For the computation of the suitable step size  at each iteration, we apply the backtracking line search \cite{Boyd2004}. Note that    $ \bq_{l}= \pdv{\mathrm{\overline{SR}}_{\mathrm{PGP}}}{\bs_{l}^{*}}=\frac{\bar{K}}{\ln2}\sum_{g=1}^{G}  \pdv{	\bar{\gamma}_{g,\mathrm{PGP}}}{\bs_{l}^{*}} $  is provided by Proposition \ref{Prop:optimPhase} below. The  projection problem $ \min_{|\phi_{n} |=1, n=1,\ldots,N}\|\bs-\tilde{\bs}\|^{2} $ under the unit-modulus constraint provides the solution of the problem described by  \eqref{sol2}. 
	In Algorithm \ref{Algoa1}, we present the outline of this procedure.
	\begin{algorithm}
		\caption{Projected Gradient Ascent Algorithm for the IRS Design}
		1.				 \textbf{Initialisation}: $ \bs_{0} =\exp\left(j\pi/2\right)\one_{N}$, $ \bPhi_{0}=\diag\left(\bs_{0}\right) $, $ 	\mathrm{\overline{SR}}_{\mathrm{PGP}}^{0}=f\left(\bPhi_{0}\right) $ given by \eqref{Maximization1}; $ \epsilon>0 $\\
		2. \textbf{Iteration} $ l $: \textbf{for} $ l=0,1,\dots, $ do\\
		3. $\bq_{l}= \frac{\bar{K}}{\ln2}\sum_{g=1}^{G}  \pdv{	\bar{\gamma}_{g,\mathrm{PGP}}}{\bs_{l}^{*}}  $, where $\pdv{	\bar{\gamma}_{g,\mathrm{PGP}}}{\bs_{l}^{*}} $ is given by Proposition \ref{Prop:optimPhase};\\
		4. \textbf{Find} $ \mu $ by backtrack line search$( f\left(\bPhi_{0}\right),\bq_{l},\bs_{l})$ \cite{Boyd2004};\\
		5. $ \tilde{\bs}_{l+1}=\bs_{l}+\mu \bq_{l} $;\\
		6. 	$ \bs_{l+1}=\exp\left(j \arg \left(\tilde{\bs}_{l+1}\right)\right) $; $ \bPhi_{l+1}= \diag\left(\bs_{l+1}\right) $;\\
		7. $ \mathrm{\overline{SR}}_{\mathrm{PGP}}^{l+1}=f\left(\bPhi_{l+1}\right) $;\\
		8. \textbf{Until} $ \| \mathrm{\overline{SR}}_{\mathrm{PGP}}^{l+1}- \mathrm{\overline{SR}}_{\mathrm{PGP}}^{l}\|^{2} <\epsilon$; \textbf{Obtain} $ \bPhi^{\star}=\bPhi_{l+1}$;\\
		9. \textbf{end for}\label{Algoa1}
	\end{algorithm}

	\begin{proposition}\label{Prop:optimPhase}
		The derivative of $ 	\bar{\gamma}_{g_{k},\mathrm{PGP}} $ with respect to $ \bs_{l}^{*} $ is provided by
		\begin{align}
			\pdv{\bar{\gamma}_{g,\mathrm{PGP}} }{\bs_{l}^{*}}=\frac{\pdv{S_{g}}{\bs_{l}^{*}}I_{g}-S_{g}\pdv{I_{g}}{\bs_{l}^{*}}}{I_{g}^{2}},\label{gam01}
		\end{align}
		where
		\begin{align}
		&\pdv{S_{g}}{\bs_{l}^{*}}=2\frac{P_{g}}{\bar{K}}\left(1-\tau_{g}^{2}\right)\bar{\delta}_{g}\bar{\delta}_{g}', \label{sg4}\\
		&\pdv{I_{k}}{\bs_{l}^{*}}=\bar{\bY}_{gg}'\!\left(\!1\!-\!\tau_{g}^{2}\!\left(1-\left(1+\bar{\delta}_{g}\right)^{2}\right)\!\right)\!+2\bar{\bY}_{gg}\tau_{g}^{2}\left(1+\bar{\delta}_{g}\right)\bar{\delta}_{g}'\nn\\
			&+\sum_{l\ne g}\left(\bar{\lambda}_{l}'\bar{\bY}_{gl}+\bar{\lambda}_{l}\bar{\bY}_{gl}'\right)\frac{\left(1+\bar{\delta}_{g}\right)^{2}}{\bar{\lambda}_{g}}\nn\\
			&+\left(1+\sum_{l\ne g}\bar{\lambda}_{l}^{2}\bar{\bY}_{gl}\right)\frac{\left(1+\bar{\delta}_{g}\right)\left(2\bar{\delta}_{g}'\bar{\lambda}_{g}-\left(1+\bar{\delta}_{g}\right)\bar{\lambda}_{g}'\right)}{\bar{\lambda}_{g}^{2}}
		\end{align}
		with
		\begin{align}
			&\bar{\delta}_{g}'=\frac{1}{\bar{b}}\tr\left(\bar{\bR}_{g}'\bT_{g}+\bar{\bR}_{g}\bT_{g}'\right),~
				\bT_{g}'=-\bT_{g}(\bT_{g}^{-1})'\bT_{g},\\
			&(\bT_{g}^{-1})'= \frac{K}{\bar{b}}\frac{\bar{\bR}_{g}'\left(1+\bar{\delta}_{g}\right)-\bar{\bR}_{g}\bar{\delta}_{g}'}{(1+\bar{\delta}_{g})^{2}},	\label{d1}\\
			&	\bar{\bY}_{gg}' =\frac{P_{g}}{\bar{b}}\left(1-\frac{1}{\bar{K}}\right)\frac{\bar{m}_{gg}'\left(1+\bar{\delta}_{g}\right)-2\bar{m}_{gg}\bar{\delta}_{g}'}{\left(1+\bar{\delta}_{g}\right)^{3}},~\\
						&		\bar{\bY}_{gl}' =\frac{1}{\bar{b}}\frac{P_{l}}{\bar{K}}	\frac{\bar{m}_{gl}'\left(1+\bar{\delta}_{l}\right)-2\bar{m}_{gl}\bar{\delta}_{l}'}{\left(1+\bar{\delta}_{l}\right)^{3}},\\
						&					\bar{\lambda}_{g}'\!=\!\frac{\rho_{g}}{\bar{b}}P_{g}	\frac{2 m_{g}\bar{\delta}_{g}'\!-\!m_{g}'\left(1\!+\!\bar{\delta}_{g}\right)}{\left(1+\bar{\delta}_{g}\right)^{3}\bar{\Psi}_{g}^{2}},~\\
						&			\!\bar{m}_{g}'\!=\!f_{g}'\!\left(\bB_{g}^{\H}\bB_{g}\right)\!,	\bar{m}_{gg}'=f_{g}'\!\left(\bar{\bR}_{g}\right)\!, 	\bar{m}_{gl}'=f_{l}'\!\left(\bB_{l}^{\H}\bar{\bR}_{g}\bB_{l}\right)\!,\label{mg12}
				\end{align}
		with 
		\begin{align}
			&f_{g}'\left(\bA\right)=\frac{\frac{\bar{K}}{\bar{b}}\left(1+\bar{\delta}_{g}\right)\left(q'(\bA)\left(1+\bar{\delta}_{g}\right)+2q(\bA)\bar{\delta}_{g}'\right)}{\left(b\left(1+\bar{\delta}_{g}\right)^{2}-{\frac{\bar{K}}{\bar{b}}q(\bar{\bR}_{g})}\right)}\nn\\
			&+\frac{\frac{\bar{K}}{\bar{b}}\left(1+\bar{\delta}_{g}\right)^{2} q(\bA)\left(2 b\left(1+\bar{\delta}_{g}\right)\bar{\delta}_{g}'-\frac{\bar{K}}{\bar{b}}q'(\bar{\bR}_{g})\right)	}{\left(b\left(1+\bar{\delta}_{g}\right)^{2}-\frac{\bar{K}}{\bar{b}}q(\bar{\bR}_{g})\right)^{2}},\label{fa}
		\end{align}
		and  $ q'\!(\bC) $ for $\bC\!=\!\bA, \bar{\bR}_{g}  $  written as $ 	q'\!(\bC)\! =\!	\tr\!\left(\bar{\bR}_{g}'\bT_{g}\bC\bT_{g}\!+\!\bar{\bR}_{g}\bT_{g}'\bC\bT_{g}\!+\!\bar{\bR}_{g}\bT_{g}\bC'\bT_{g}\!+\!\bar{\bR}_{g}\bT_{g}\bC\bT_{g}'\right)\!, $
		where each term is computed by  Lemma \ref{lem1}. The expressions of  $\bar{ \delta}_{g} $, $ \Psi_{g} $, $ \bar{\bY}_{gg} $, and $ \bar{\bY}_{gl} $, given in Theorem \ref{theorem:PGP} and Corollary \ref{cor1} together with   $  	\bar{\delta}_{g}'$, $  \bar{\lambda}_{g}' $, $  \bar{\bY}_{gg}' $, $  \bar{\bY}_{gl}' $, $ 	\bar{m}_{g}' $, 	$ \bar{m}_{gg}' $, and $ 	\bar{m}_{gl}' $  are obtained by solving the system of fixed-point equations \eqref{s1}-\eqref{s6}, \eqref{d1}-\eqref{mg12}.
	\end{proposition}
	\proof The proof  is given in Appendix~\ref{optimPhase}.\endproof
	
The RBM beamforming design is based on the gradient ascent and offers a significant advantage  because  the gradient ascent is derived in a closed-form. Note that it has low computational complexity because it consists of simple matrix operations. Specifically,  the complexity of (35) is $ \mathcal{O}\left(G(MN^{2}+N+M)\right) $. Obviously, the derivative is a function of the fundamental system parameters  $ G $, $ M $, and $N $ with the number of IRS elements having the higher (square) impact.

	\subsection{Power Allocation Optimization}
	Now, for a fixed RBM $ \bPhi $, the objective is the optimization over $ \bp $. In particular, we have 
	\begin{align}\begin{split}
			(\mathcal{P}4)~~~~~~~\max_{\bp\ge 0} ~~~	&		\mathrm{\overline{SR}}_{\mathrm{PGP}}\\
			\mathrm{s.t}~~\;\!&\sum_{g=1}^{G}P_{g}\le P_{\mathrm{max}},
		\end{split}\label{Maximization} 
	\end{align}
	where  	$\mathrm{\overline{SR}}_{\mathrm{PGP}} $ is given by \eqref{Maximization1}. 	This problem is not convex but a local optimal solution can be obtained by using a weighted minimum mean square error (WMMSE) reformulation of the sum SE maximization.  By denoting $\bc_{g}\!=\![c_{g1}, \ldots,c_{gG} ]^{\T}  $,  the SINR $ \bar{\gamma}_{g,\mathrm{PGP}} $ can be expressed as a function of the downlink power coefficients given by the vector $ \bp $ as
	\begin{align}
		\bar{\gamma}_{g,\mathrm{PGP}} =\frac{P_{g} q_{g}}{\bc_{g}^{\T}\bp+t_{g}^{2}},
	\end{align}
	where
	\begin{align}
		q_{g}&\!=\!\frac{1}{\bar{K}}\!\left(\!1\!-\!\tau^{2}\right)\!\bar{\delta}_{g}^{2},\! ~	t_{g}^{2}\!=\!\frac{\left(1+\bar{\delta}_{g}\right)^{2}}{\bar{\lambda}_{g}},\\
		c_{gg}&\!=\!\frac{1}{\bar{b}}\!\left(\!1\!-\!\frac{1}{\bar{K}}\!\right)\!	\frac{\bar{m}_{gg}}{\left(1\!+\!\bar{\delta}_{g}\right)^{2}}	\!\left(\!1-\tau^{2}\!\left(\!1\!-\!\left(1+\bar{\delta}_{g}^{2}\right)^{2}\right)\!\right)\!,~\forall g \\
		c_{gi}&= \frac{1}{\bar{b}}	\frac{1}{\bar{K}}\frac{m_{gl}}{\left(1+\bar{\delta}_{l}\right)^{2}}\frac{\bar{\lambda}_{l}\left(1+\bar{\delta}_{g}\right)^{2}}{\bar{\lambda}_{g}},~\forall g, ~\forall i\ne g.
	\end{align}
	
	Now, the optimization problem  becomes
	\begin{align}\begin{split}
			(\mathcal{P}5)~~~~~~~\max_{\bp\ge 0} ~~~	&		\bar{K}\sum_{g=1}^{G}\log_{2}(1 + 	\frac{P_{g} q_{g}}{\bc_{g}^{\T}\bp+t_{g}^{2}})\\
			\mathrm{s.t}~~\;\!&\sum_{g=1}^{G}P_{g}\le P_{\mathrm{max}}.
		\end{split}\label{Maximization9} 
	\end{align}

 For the MMSE reformulation, we assume the  single‐input and single‐output (SISO) channel that corresponds to this SINR given by
	\begin{align}
		\tilde{y}_{g}=\sqrt{P_{g} q_{g}}s_{g}+\sum_{i=1}^{G}\sqrt{P_{g}c_{gi}}s_{i}+n_{g},
	\end{align}
	where $ \tilde{y}_{g} $ is the received signal, $ s_{g }\in \mathbb{C} $ denotes the normalized and independent random data signal with $ \EE[|s_{g}|^{2}=1] $, and $ n_{g}\sim \cC\cN\left( 0, t_{g}^{2} \right) $. In such case, the receiver can compute an estimate $\hat{s}_{g}=v_{g}^{*} 	\tilde{y}_{g} $ of the desired signal $ s_{g} $, where  $ v_{g} $ as a scalar combining coefficient. The resulting MSE 	$ e_{g}(\bp,v_{g}) =[|\hat{s}_{g}-s_{g}|^{2}]$ is written as
	\begin{align}
		e_{g}(\bp,v_{g})=v_{g}^{2}\left(P_{g}q_{g}+\bc_{g}^{\T}\bp+t_{g}^{2}\right)-2 v_{g}\sqrt{q_{g}P_g}+1.\label{mse1}
	\end{align}
	
	The coefficient $ 	v_{g} $, minimizing  the MSE 	$ e_{g}(\bp,v_{g})$ for a given $ \bp $, becomes 
	\begin{align}
		v_{g}=\frac{\sqrt{P_{g} q_{g}}}{P_{g}q_{g}+\sum_{i=1}^{G}P_{g}c_{gi}+t_{g}^{2}}.
	\end{align}
	
	Plugging $ v_{g} $ into \eqref{mse1}, $ e_{g} $ becomes $ 1/\left(1+	\bar{\gamma}_{g,\mathrm{PGP}} \right) $. Based on the  weighted MMSE method, we introduce the auxiliary weight $ d_{g}\ge 0 $ for the MSE $  e_{g}$ and solve the following  problem
	\begin{align}\begin{split}
			(\mathcal{P}6)~\min_{\substack{\bp\ge 0,\\
					\{v_{g}, d_{g}\ge 0: g=1,\ldots,G\}}} 	&		\bar{K}\!\sum_{g=1}^{G}\!d_{g} e_{g}(\bp,\bv_{g})-\ln(d_{g})\\
			\mathrm{s.t}~~\;\!&\sum_{g=1}^{G}P_{g}\le P_{\mathrm{max}}.
		\end{split}\label{Maximization10} 
	\end{align}
	
	Note that both $ 	(\mathcal{P}5) $ and $ 	(\mathcal{P}6) $ are equivalent, meaning that  they have the same  optimal solution. The equivalence  stems from the fact that the optimal $ d_{g} $ in \eqref{Maximization10} is $ 1/ e_{g}=(1+	\bar{\gamma}_{g,\mathrm{PGP}})$. The advantage of  the reformulation results in the following lemma  adapted from \cite[Th. 3]{Shi2011}.
	
	\begin{lemma}
		The block descent coordinate algorithm, provided in Algorithm \ref{Algoa2}, converges to a local optimum of $ 	(\mathcal{P}6) $ in terms of AO among three blocks of variables $ \{v_{g}:  g=1,\ldots,G\} $, $ \{d_{g}:  g=1,\ldots,G\} $, and $ \bp $.
	\end{lemma}

	\begin{algorithm}
		\caption{Block coordinate descent algorithm for solving $ 	(\mathcal{P}6) $}
		1.				 \textbf{Initialisation}: Set $ \bp=\frac{P_{\mathrm{max}}}{G}\one_{G} $ and the solution accuracy $ \epsilon >0 $, \\
		2. \textbf{while} the objective function in \eqref{Maximization10} is not improved more than $ \epsilon $ \textbf{do}\\
		3. $ 	v_{g}=\frac{\sqrt{P_{g} q_{g}}}{P_{g}q_{g}+\sum_{i=1}^{G}P_{g}c_{gi}+t_{g}^{2}}., $ $ g=1,\ldots, G $\\
		4. $ d_{g}=1/e_{g}(\bp,v_{g}), $ $ g=1,\ldots, G $\\
		5. Solve the following problem for the current values of $ v_{g} $ and $ d_{g} $:\\\vskip-12mm
		\begin{align}\begin{split}
				(\mathcal{P}7)~~~~~~~\min_{\bp \ge 0} ~~~	&		\bar{K}\sum_{g=1}^{G}d_{g} e_{g}(\bp,\bv_{g})\\
				\mathrm{s.t}~~\;\!&\sum_{g=1}^{G}P_{g}\le P_{\mathrm{max}},\\
			\end{split}\label{Maximization11} 
		\end{align}	\vskip-2mm
		6. Update $ \bp $ by the obtained solution to \eqref{Maximization11}\\
		7. \textbf{end while}\\
		8. \textbf{Output:} $ \bp^{\star} $\label{Algoa2}
	\end{algorithm}
	
	Algorithm \ref{Algoa2}  describes the whole procedure for the power optimization and Step $ 5 $  includes a subproblem that needs to be solved in every iteration. It  can be solved in closed-form as
	\begin{align}
		P_{g}=\min\left(P_{\mathrm{max}},\frac{q_{g}d_{g}^{2}v_{g}^{2}}{\left(q_{g}d_{g}v_{g}^{2}+\sum_{i=1}^{G}d_{i}v_{i}^{2}c_{ig}\right)^{2}}\right),
	\end{align}
where  $ \sqrt{P_{g}}, g=1\ldots, G $ are treated as optimization variables. Basically, the problem is decomposed into $ G $ independent subproblems, where each one of them concerns a quadratic minimization under a bound constraint. 

	The power allocation presents a similar complexity to the RBM design since Algorithm 2 consists of similar  matrix operations, i.e., its complexity is  $ \mathcal{O}\left(G(MN^{2}+N+M)\right) $.
	
	\begin{remark}
		Both Algorithms \ref{Algoa1} and \ref{Algoa2} converge quickly and have low computation complexity. Also, since they achieve to obtain just a local optimum, different initializations are expected to lead to different solutions of the overall algorithm.
	\end{remark}

	\section{Numerical Results}\label{Numerical}
	In this section, we elaborate on the numerical results of the sum SE in IRS-assisted systems under the FDD protocol by applying PGP of the JSDM method. MC simulations in terms of $ 10^{3} $ independent channel realizations verify the deterministic equivalent analysis. 
	
	The simulation setup includes a uniform linear array (ULA) of $ M =100$ antennas  at the BS, which serves  $ K = 30 $ UEs while aided by an
	IRS with a uniform planar array (UPA) of $ N=100 $ elements. UEs are equally shared among $ G=6 $ groups, which means that each group consists of $ 5 $  UEs.  The correlation matrices $ \bR_{\mathrm{BS},g_{k}} $ and $ \bR_{\mathrm{IRS},g_{k}} $ are  obtained similar to \cite{Hoydis2013} and \cite{Bjoernson2020}, respectively.  Note that the size of each IRS element is given by $ d_{\mathrm{H}}\!=\!d_{\mathrm{V}}\!=\!\lambda/4 $. Furthermore, the  path-losses for the BS-to-IRS and IRS-to-UE $ k $ links are given by \cite{Wu2019,Kammoun2020}
	\begin{align}
		\beta_{1}=\frac{C_{1}}{d_{1}^{\al_{1}}},~~~\beta_{2,k}=\frac{C_{2}}{d_{2,k}^{\al_{2}}}  ,
	\end{align}
	where $ \al_{1} $ and $ d_{1} $ are the path-loss exponent and distance concerning the former link  while $ \al_{2} $ and $ d_{2,k} $ are the path-loss exponent and distance concerning the latter link. In the case of $ \beta_{\mathrm{d},k} $, we assume the same parameters as for $ \beta_{2,k} $, and additionally, we  consider a penetration loss of $ 15~\mathrm{dB} $. The choice of these values  relies on the 3GPP Urban Micro (UMi) scenario from TR36.814 for a carrier frequency of $ 2.5 $ GHz and a noise level of $ -80 $ dBm. Especially, the path-losses for  $ \bH_{1} $ and $ \bh_{2,k} $ are generated according to the LOS and NLOS versions \cite{Access2010}. Also, $ C_{1}=26 $ dB and $ C_{2}=28 $ dB, which are  the path-losses at a reference distance
	of $ 1$m
while each UE includes a single
	0dBi antenna \cite{Bjoernson2019b}.  Also, we have chosen $ \bar{b}=12 $ and $ \bar{r}=12 $. The latter value is justified because under these setting the rank of the channel covariance matrix   $ \bR_{g} $ is $\bar{r}=27  $ while only $ 12 $ of them are important. Unless otherwise stated, this set of parameters values is used during the simulations.
	  Note that $ \sigma^2=-174+10\log_{10}B_{\mathrm{c}} $, where $B_{\mathrm{c}}=200~\mathrm{kHz}$. 
	
\begin{figure}%
		\centering
		\subfigure[]{	\includegraphics[width=0.9\linewidth]{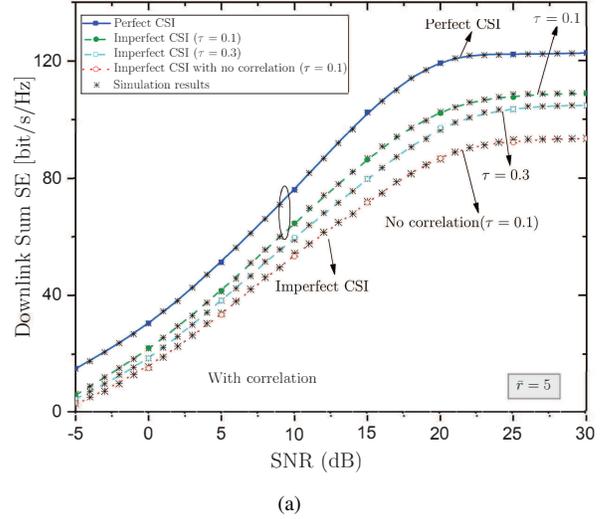}}\qquad
		\subfigure[]{	\includegraphics[width=0.9\linewidth]{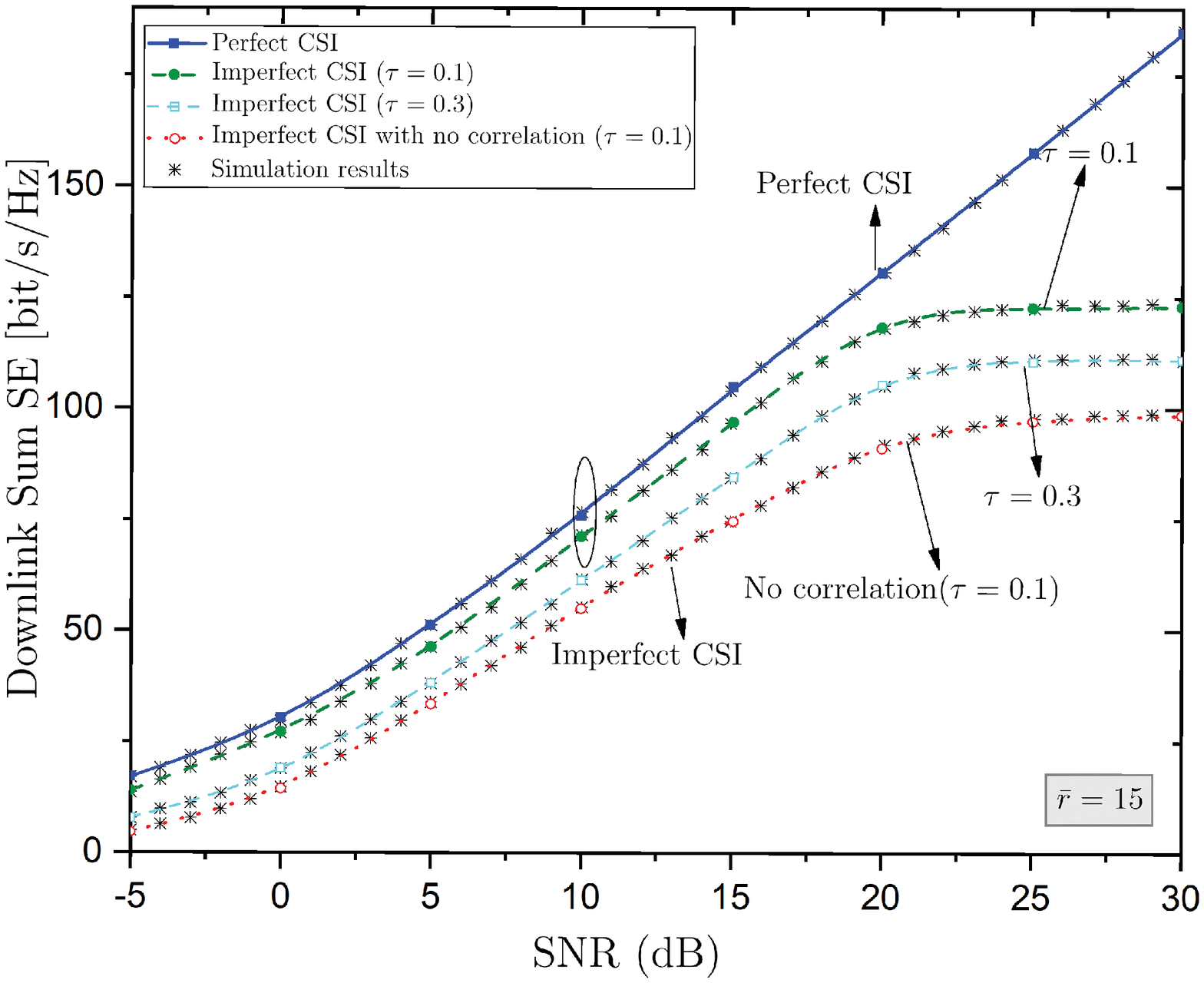}}\\
		\caption{Sum SE of an IRS-assisted MIMO system with FDD in the cases of perfect/imperfect CSI and correlated/independent Rayleigh fading  versus the SNR: (a) $\bar{r}=5$ ; (b) $ \bar{r}=12 $.}
		\label{Fig2}
	\end{figure}

	Figs. \ref{Fig2}.(a) and \ref{Fig2}.(b) illustrate the impact of the effective rank of the channel covariance matrix of each group on the   performance of JSDM in IRS-assisted systems in the cases of perfect and imperfect CSI.  
	Under perfect CSI conditions ("solid" lines) with $ \tau=0 $, we notice that in the case of choosing $ \bar{r}=5 $ (Fig. \ref{Fig2}.(a)), PGP saturates at high SNR since this effective rank value is too small. In particular, such a choice does not allow for consideration of the substantial eigenmodes by the pre-beamforming matrix and results in large inter-group interference.  On the contrary, a better (higher) choice for $ \bar{r}$, i.e., $ \bar{r}=15$ (see Fig. \ref{Fig2}.(b)) allows the inclusion of the significant eigenmodes and no interference-limited behaviour is noticed until $ 30 $ dB. Of course, the rate saturates at some larger SNR, belonging outside the scope of practical applications. Obviously, the choice of the rank value depends on the channel covariance matrix. In the case of imperfect CSI ($ \tau=0.1 $), met in practice,  saturation is observed in both figures ("dashed" lines). However, when we have $ \bar{r}=5 $, the saturation starts much earlier ($ 25 $ dB) while a choice of $ \bar{r}=15 $ (Fig. \ref{Fig2}.(b)) is more robust regarding the performance since the saturation is encountered after $ 21.7 $ dB. Similar observations hold when $ \tau=0.3 $. The inference from these two figures is that a good selection  of $ \bar{r} $, where the major eigenmodes are taken into account, defines the performance. A small value results in severe performance degradation while a value close to the full rank $\bar{r}$ will result in a dimensionality bottleneck while not achieving any profit concerning the interference. In these figures, we also show the degradation of the sum SE when no correlation ("dotted" lines) is taken into account because in such a case the IRS cannot be optimized.
	
	To elaborate further on the saving advantage of the JSDM method, we would like to highlight that after the downlink training of all groups ($ G=6 $) with effective channel dimension $ \bar{b}=12 $, each of the $ \bar{K} $ UEs in each group has to feedback its $ 12\times 1$ channel vector, which equals to $ 5\times 6\times 12=360 $ quantized complex channel coefficients.\footnote{This conclusion assumes that downlink training has  already taken place at some previous stage. Note that this work does not perform downlink training but this task and its effects are left for future study.} However, in this case, where the BS is deployed with $ M=100 $ antenna aimed at serving $ K=30 $ UEs in total, $ 100 $ orthogonal downlink training symbols are required  while a total of $ 30\times 100 $ channel coefficients will be fed back, which means the achievement of saving by a factor of $ 10 $. 

\begin{figure}[!h]
		\begin{center}
			\includegraphics[width=0.9\linewidth]{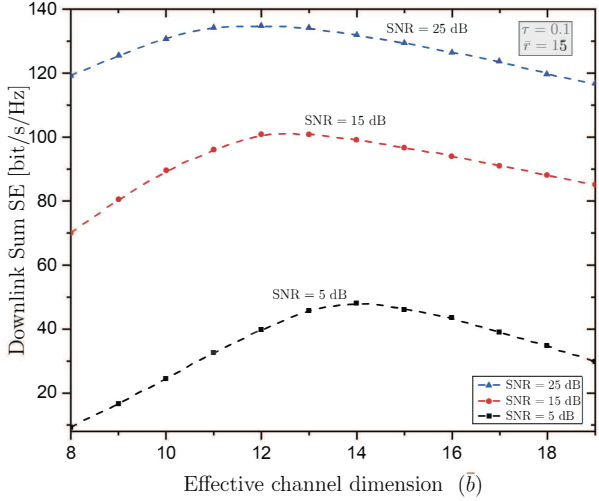}
			\caption{Sum SE of an IRS-assisted MIMO system with FDD in the cases of imperfect CSI ($ \tau=0.1 $) versus the 	effective channel dimension $\bar{b}$ for varying $ \mathrm{SNR} $ values.}
			\label{Fig3}
		\end{center}
	\end{figure}

	In Fig. \ref{Fig3}, we depict the impact of the parameter $ \bar{b} $ for a specific $ \bar{K} $ and when $ \tau=0.1 $ (imperfect CSI) for varying SNR values, being $ \mathrm{SNR}=5, 15, 25 $ dB. Specifically, after selecting $ \bar{r}=15 $ according to the  discussion in the previous paragraph and subject to the constraint $ \bar{K}\le \bar{b}\le M- \bar{r}\left(G-1\right) $, we observe that the sum SE does not increase monotonically but there is an optimal value $ \bar{b} $ because its increase results in a trade-off between a larger dimensionality overhead and a better channel conditioning. Moreover, we observe that the higher the SNR, the smaller the optimal $ \bar{b} $ becomes. Also, we notice that at higher SNR,  the  range of sum SE values takes higher values but the impact of $ \bar{b} $ variation is smaller  with increasing SNR as witnessed by the slopes of the corresponding lines.
	
\begin{figure}[!h]
		\begin{center}
			\includegraphics[width=0.9\linewidth]{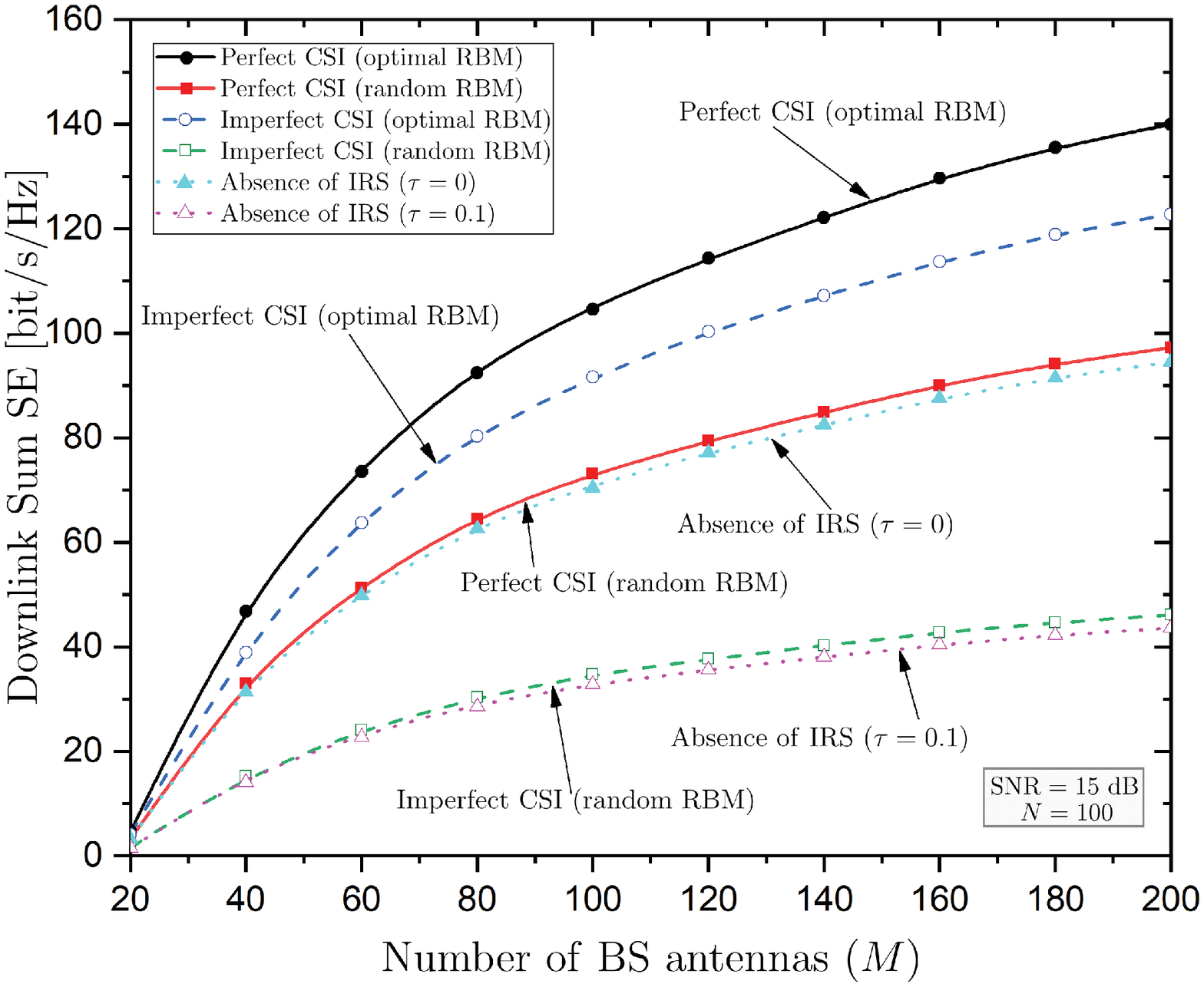}
			\caption{Sum SE of an IRS-assisted MIMO system with FDD versus the number of BS antennas $ M $ in the cases of i) perfect/imperfect CSI ($ \tau=0.1 $) with optimal and random RBM, ii) without IRS and perfect/imperfect CSI.}
			\label{Fig4}
		\end{center}
	\end{figure}

		Fig. \ref{Fig4} shows  the achievable sum SE versus the number of BS antennas $ M $ for both perfect and imperfect CSI, when $ N=100 $ elements. We observe that  $ \mathrm{\overline{SR}}_{\mathrm{PGP}} $ exhibits a similar dependence on $ M $ as  on $ N $ in the previous figure. Thus, when $ M $ grows large, the sum SE increases without limit in both cases of perfect and imperfect CSI.  In general, these two figures indicate that an IRS-assisted system performs better with larger values of IRS elements and BS antennas. In particular,   the latter is further interesting because it agrees with the massive MIMO design trend that has already started to be  implemented. In addition, we verify that the performance increases in the presence of the IRS.

\begin{figure}[!h]
		\begin{center}
			\includegraphics[width=0.9\linewidth]{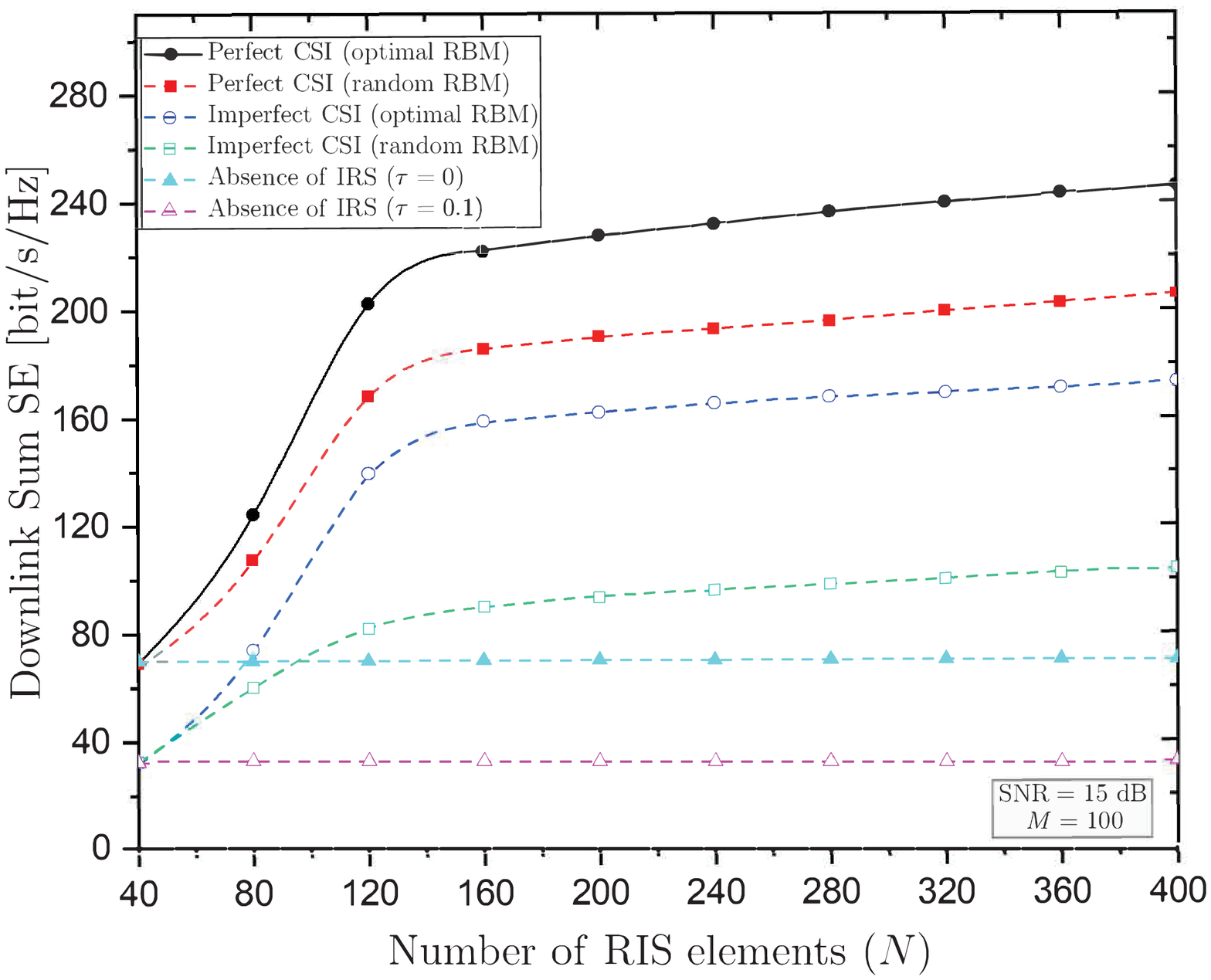}
			\caption{Sum SE of an IRS-assisted MIMO system with FDD versus the number of IRS elements $ N $  in the cases of i) perfect/imperfect CSI ($ \tau=0.1 $) with optimal and random RBM, ii) without IRS and perfect/imperfect CSI.}
			\label{Fig41}
		\end{center}
	\end{figure}

	Fig. \ref{Fig41} illustrates the achievable sum SE versus the number of IRS elements $ N $ for both cases of perfect and imperfect CSI, when $ M=100 $ antennas.  We notice that the sum SE increases with the increasing number of IRS elements in both cases, but the increase is slower in the case of imperfect CSI. Also, disregarding the IRS size, defined by its elements, it always provides better performance compared to its absence. In other words, its application is beneficial because it results in a further enhancement of the channel additionally to the direct signal. Moreover, we consider the scenario of "random" phase shifts and we show that the RBM optimization enhances significantly the performance.

	
		\begin{figure}[!h]
		\begin{center}
			\includegraphics[width=0.9\linewidth]{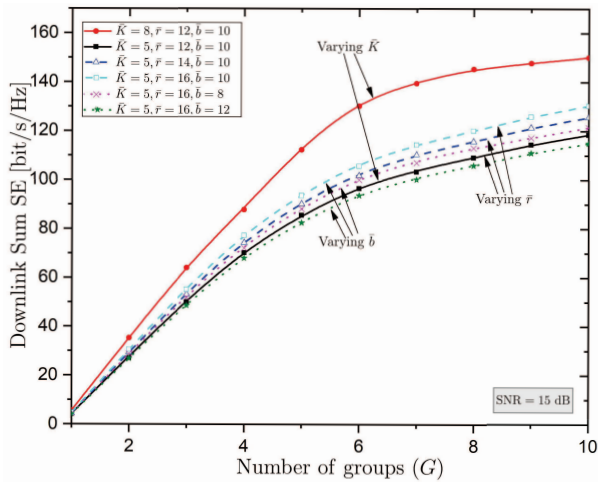}
			\caption{Sum SE of an IRS-assisted MIMO system with FDD  versus the number of groups $ G $ in the case of imperfect CSI ($ \tau=0.1 $) for optimal and random $ \mathrm{RBM} $ for varying effective rank $ \bar{r} $, 	 effective channel dimension $ \bar{b} $, and UEs per group $ \bar{K} $.}
			\label{Fig5}
		\end{center}
	\end{figure}

	Fig~ \ref{Fig5} depicts the sum SE versus the number of groups  in the case of imperfect CSI ($ \tau=0.1 $) while varying  
	the effective rank $ \bar{r} $, 	the effective channel dimension $ \bar{b} $, and UEs per group $ \bar{K} $.  We notice that the sum SE increases with the number of groups but this increase becomes slower when the number of UEs per group grows due to increased multiuser interference while the rest of the parameters are the same. In addition, for a specific $ \bar{K} $, an increase of the channel rank results in an increase of the sum SE but caution should be taken since the overhead (dimensionality) increases too. Moreover, by increasing $ \bar{b} $, the sum SE increases until a specific $ \bar{b} $  equal to $ 10 $ while keeping increasing $ \bar{b} $ results in lower sum SE due to larger dimensionality cost 
	 again.
	\begin{figure}%
	\centering
	\subfigure[]{	\includegraphics[width=0.9\linewidth]{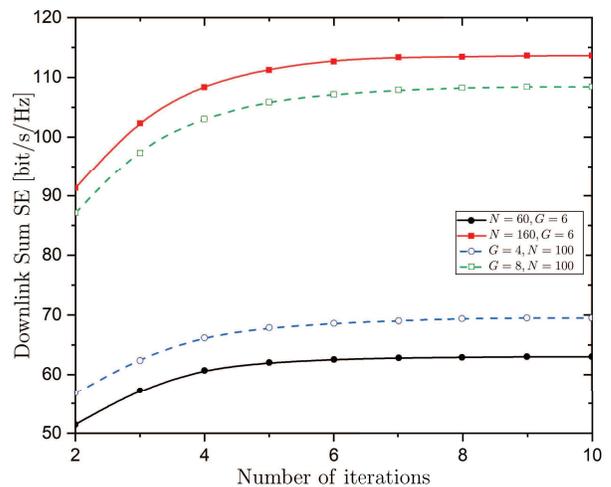}}\qquad
	\subfigure[]{	\includegraphics[width=0.9\linewidth]{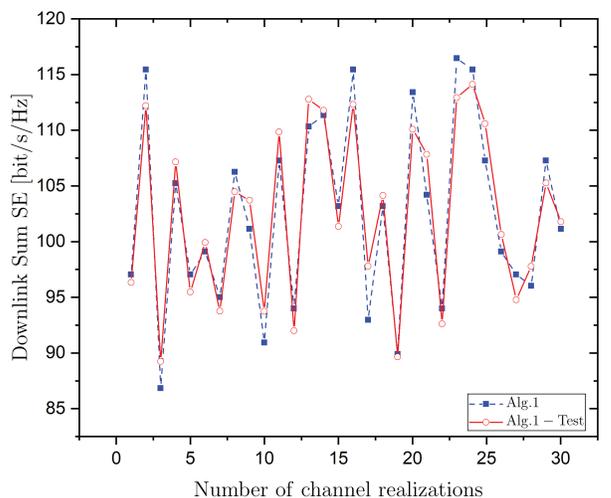}}\\
	\caption{Sum SE of an IRS-assisted MIMO system with FDD versus: (a) the number of iterations  for varying IRS elements $ N $ and groups $ G $; (b) $ 30 $ channel realizations.}
	\label{Fig6}
\end{figure}

	Fig. \ref{Fig6}.(a) illustrates the convergence of the proposed overall algorithm based on AO, which consists of the subproblems of RBM and transmit power optimizations. Specifically, we have depicted the   sum SE versus the number of iterations for varying numbers of IRS elements and UE groups. As can be seen, the convergence of the optimization is fast in all cases, where the algorithm has converged at most in $ 8 $ iterations (see the "dashed-square" line). Also, as expected, as the number of IRS elements and groups increases, more iterations are needed until convergence since the number of optimization variables has increased.

	The non-convexity of the overall optimization problem suggests that its solution depends on the initial point, i.e., different initial points result in different locally optimal solutions. Fig. \ref{Fig6}.(b) investigates this dependence on the initializations by accounting for $ 30 $ channel realizations. The initialization of the overall algorithm including Algs. $ 1 $ and $ 2 $ assumes that $ \bs_{0} =\exp\left(j\pi/2\right)\one_{N}$ and $ \bp=\frac{P_{\mathrm{max}}}{G}\one_{G}  $ as mentioned previously. "Alg. 1-Test" in the figure assumes the best initial point out of $ 100 $ random initial points for each channel instance. We observe indeed that different initializations result in different solutions and that the sum SE in both cases is almost the same, which means that this  phase shifts selection for initialization is a good choice. 
	
	\section{Conclusion} \label{Conclusion} 
	In this paper, we tackled the channel non-reciprocity witnessed in IRS-assisted communication systems. For this reason, we assumed an mMIMO system assisted with an IRS operating in FDD. To this end, we applied the JSDM method by grouping  the UEs with the same covariance matrix to improve the performance, which, otherwise, would be afflicted due to prohibitive feedback overhead as the number of BS antennas and IRS elements increases. Based on DE tools and S-CSI knowledge, we derived the sum SE accounting for correlated Rayleigh fading. Next, we formulated the optimization  problem maximizing the sum SE under RBM and power budget constraints. To solve the nonconvex optimization, we proposed an efficient AO algorithm that can be performed at every several coherence intervals, which results in further reduction of the feedback overhead and computational complexity. Among the observations,  we would like to highlight that the rank of the covariance matrix is a crucial parameter having a direct impact on the performance. Its increase improves the performance due to better channel conditioning while it burdens the feedback overhead. 
 Other interesting directions could be the  extension of this work to account for 3-dimensional beamforming  with multiple beams in the elevation angle and the consideration of multiple distributed IRS.
	
	\begin{appendices}
		\section{Proof of Theorem~\ref{theorem:PGP}}\label{theorem1}
		The derivation of the DE SINR $ \bar{\gamma}_{g_{k},\mathrm{PGP}} $ is obtained by following the approach in \cite{Wagner2012} but with extension to imperfect CSI that brings considerable differences. Hence, taking into account for imperfect CSI, the proof consists of the derivation of the DEs of four parts: i) The power normalization term $ \Psi_{g} $, ii) the  desired signal power part $ 	\mathrm{DS}_{g_{k}} $, iii) the self-group interference power part $ \mathrm{SGI}_{g_{k}} $, and iv) the inter-group interference power part $ \mathrm{IGI}_{g_{k}} $.
		
		Regarding  $ \Psi_{g} $, we have
		\begin{align}
			\Psi_{g} &=\sum_{k=1}^{\bar{K}}p_{g_{k}}\hat{\bar{\bh}}_{g_{k}}^{\H}{\bSigma}_{g}^{-2}\hat{\bar{\bh}}_{g_{k}}\\
			&=\frac{1}{\bar{b}}\sum_{k=1}^{\bar{K}}p_{g_{k}}\frac{\frac{1}{\bar{b}}\hat{{\bh}}_{g_{k}}^{\H}\bB_{g}\bSigma_{g_{[k]}}^{-2}\bB_{g}^{\H}\hat{{\bh}}_{g_{k}}}{\left(1+\frac{1}{\bar{b}}\hat{{\bh}}_{g_{k}}^{\H}\bB_{g}\bSigma_{g_{[k]}}^{-1}\bB_{g}^{\H}\hat{{\bh}}_{g_{k}}\right)^{\!2}}\label{psi1}\\
			&\asymp \frac{1}{\bar{b}}\sum_{k=1}^{\bar{K}}p_{g_{k}}\frac{\frac{1}{\bar{b}}\tr\left(\bar{\bR}_{g}\bSigma_{g_{[k]}}^{-2}\right)}{\left(1+\frac{1}{\bar{b}}\tr\left(\bar{\bR}_{g}\bSigma_{g_{[k]}}^{-1}\right)\right)^{\!2}}\label{psi2}\\
			&\asymp \frac{1}{\bar{b}}P_{g}\frac{\bar{m}_{g}}{\left(1+\bar{\delta}_{g}\right)^{\!2}},\label{psi3}
		\end{align}
		where $ \bSigma_{g_{[k]}}=\frac{1}{\bar{b}}\hat{\bar{\bH}}_{g_{[k]}}\hat{\bar{\bH}}^{\H}_{g_{[k]}}+ \al  \Id_{\bar{b}} $. In \eqref{psi1}, we applied the matrix inversion lemma twice \cite[Lem. 1]{Hoydis2013} and, in \eqref{psi2}, we  used \cite[Lem. 4]{Hoydis2013}. The last step makes use of \cite[Th. 1]{Hoydis2013} with $ \bar{m}_{g} $ and $ \bar{\delta}_{g} $ defined in Theorem \ref{theorem:PGP}. Note that $ \bar{m}_{g} $ is basically the derivative of  $ \bar{\delta}_{g} $ with respect to $ \al $, i.e., $ \bar{m}_{g}=\bar{\delta}_{g}' =\frac{1}{\bar{b}}\tr\left(\bar{\bR}_{g}\bT_{g}'\right)$, where $ 	\bT_{g}'=\bT_{g}\left(\frac{{\bar{K}}}{\bar{b}}\frac{\bar{\bR}_{g}\bar{\delta}_{g}'}{1+\bar{\delta}_{g}}+ \Id_{\bar{b}}\right)\bT_{g} $. Its expression  is obtained after some algebraic manipulations.
		
		In the case of $ 	\mathrm{DS}_{g_{k}} $, we have
		\begin{align}
		&	\bh_{g_{k}}^{\H}\bB_{g} {\bSigma}_{g} \bB_{g}^{\H}\hat{\bar{\bh}}_{g_{k}}	=\frac{\frac{1}{\bar{b}}\bh_{g_{k}}^{\H}\bB_{g} \bSigma_{g_{[k]}}\bB_{g}^{\H} \hat{{\bh}}_{g_{k}}}{1+\frac{1}{\bar{b}}\hat{{\bh}}_{g_{k}}^{\H}\bB_{g} \bSigma_{g_{[k]}}\bB_{g}^{\H} \hat{{\bh}}_{g_{k}}}\label{ds1}\\
			&=\frac{\sqrt[]{1-\tau_{g}^2}\frac{1}{\bar{b}}\bh_{g_{k}}^{\H}\bB_{g} \bSigma_{g_{[k]}} \bB_{g}^{\H}\bh_{g_{k}}}{1+\frac{1}{\bar{b}}\hat{{\bh}}_{g_{k}}^{\H}\bB_{g} \bSigma_{g_{[k]}} \bB_{g}^{\H}\hat{{\bh}}_{g_{k}}}
			+\frac{\tau_{g}\frac{1}{\bar{b}}\bh_{g_{k}}^{\H}\bB_{g} \bSigma_{g_{[k]}}\bB_{g}^{\H} \bz_{g_{k}}}{1+\frac{1}{\bar{b}}\hat{{\bh}}_{g_{k}}^{\H}\bB_{g} \bSigma_{g_{[k]}}\bB_{g}^{\H} \hat{{\bh}}_{g_{k}}}\label{ds2}\\
			&=\frac{\sqrt[]{1-\tau_{g}^2}\bar{\delta}_{g}}{1+\bar{\delta}_{g}},\label{ds3}
		\end{align}
		where in \eqref{ds1}, we  applied the matrix inversion lemma  \cite[Lem. 1]{Hoydis2013}, and, in \eqref{ds2}, we  used \eqref{imp}. In \eqref{ds3}, we  used \cite[Lem. 4]{Hoydis2013} by considering the independence between $ \bh_{g_{k}} $ and $ \bz_{g_{k}} $. Next, we applied \cite[Lem. 4]{Hoydis2013}, \cite[Lem. 3]{Hoydis2013}, and \cite[Th. 1]{Hoydis2013}.
		
		For $ \mathrm{SGI}_{g_{k}} $, we have
		\begin{align}
			&\bh_{g_{k}}^{\H}\bB_{g} {\bSigma}_{g}\bB_{g}^{\H}\hat{{\bH}}_{g_{[k]}}\bP_{g_{[k]}}\hat{{\bH}}_{g_{[k]}}^{\H}\bB_{g}{\bSigma}_{g}\bB_{g}^{\H}\bh_{g_{k}}\nn\\
			&=\bh_{g_{k}}^{\H}\bB_{g} \bar{\bSigma}_{g_{[k]}}\bB_{g}^{\H}\hat{{\bH}}_{g_{[k]}}\bP_{g_{[k]}}\hat{{\bH}}_{g_{[k]}}^{\H}\bB_{g}{\bSigma}_{g}\bB_{g}^{\H}\bh_{g_{k}}\nn\\
			&+\bh_{g_{k}}^{\H}\bB_{g}\left( {\bSigma}_{g}\!-\!\bar{\bSigma}_{g_{[k]}}\right)\bB_{g}^{\H}\hat{\bar{\bH}}_{g_{[k]}}\bP_{g_{[k]}}\hat{\bar{\bH}}_{g_{[k]}}^{\H}\bB_{g}{\bSigma}_{g}\bB_{g}^{\H}\bh_{g_{k}}\!\label{SGI1}\\
			&=\frac{1}{b^2}\bh_{g_{k}}^{\H}\bB_{g}\bD_{g} \bB_{g}^{\H}\bh_{g_{k}}-\frac{c_{0}}{b^2}\bh_{g_{k}}^{\H}\bB_{g}\bSigma_{g} \bB_{g}^{\H}\bh_{g_{k}}\bh_{g_{k}}^{\H}\bB_{g}\bD_{g} \bB_{g}^{\H}\bh_{g_{k}}\nn\\
			&-\frac{c_{1}}{b^2}\bh_{g_{k}}^{\H}\bB_{g}\bSigma_{g} \bB_{g}^{\H}\bz_{g_{k}}\bz_{g_{k}}^{\H}\bB_{g}\bD_{g} \bB_{g}^{\H}\bh_{g_{k}}\nn\\
			&			-\frac{c_{2}}{b^2}\bh_{g_{k}}^{\H}\bB_{g}\bSigma_{g} \bB_{g}^{\H}\bh_{g_{k}}\bz_{g_{k}}^{\H}\bB_{g}\bD_{g} \bB_{g}^{\H}\bh_{g_{k}}\nn\\
			&-\frac{c_{2}}{b^2}\bh_{g_{k}}^{\H}\bB_{g}\bSigma_{g} \bB_{g}^{\H}\bz_{g_{k}}\bh_{g_{k}}^{\H}\bB_{g}\bD_{g} \bB_{g}^{\H}\bh_{g_{k}}\label{SGI2}.
		\end{align}
		In \eqref{SGI1}, we used that $ {\bSigma}_{g}-\bar{\bSigma}_{g_{[k]}}=-{\bSigma}_{g}\left({\bSigma}_{g}^{-1}-\bar{\bSigma}_{g_{[k]}}^{-1}\right)\bar{\bSigma}_{g_{[k]}} $, where $ {\bSigma}_{g}^{-1}-\bar{\bSigma}_{g_{[k]}}^{-1}=\bB_{g}^{\H}(c_{0}\bh_{g_{k}}\bh_{g_{k}}^{\H}$   $
		+c_{1}\bz_{g_{k}}\bz_{g_{k}}^{\H}+c_{2}\left(\bh_{g_{k}}\bz_{g_{k}}^{\H}+\bz_{g_{k}}\bh_{g_{k}}^{\H}\right) )\bB_{g}$ with $ c_{0}=1-\tau_{g}^2 $, $ c_{1}=\tau_{g} $, and $ c_{2}=\tau_{g}\sqrt{1-\tau_{g}^2} $. In \eqref{SGI2}, we denoted $ \bD_{g}= \bSigma_{g}\bB_{g}^{\H}\hat{\bar{\bH}}_{g_{[k]}}\bP_{g_{[k]}}\hat{\bar{\bH}}_{g_{[k]}}^{\H}\bB_{g}\bSigma_{g}$. Similar to previous derivations, we obtain
		\begin{align}
			\frac{1}{b^2}	\bh_{g_{k}}^{\H}\bB_{g}\bSigma_{g}\bB_{g}^{\H}\bh_{g_{k}}&\asymp\frac{u\left(1+c_{1}u\right)}{1+u},\\
		\frac{1}{b^2}\bh_{g_{k}}^{\H}\bB_{g}\bSigma_{g}\bB_{g}^{\H}\bz_{g_{k}}	&\asymp\frac{-c_{2}u^{2}}{1+u},\\
				\frac{1}{b^2}\bh_{g_{k}}^{\H}\bB_{g}\bD_{g}\bB_{g}^{\H}\bh_{g_{k}}&\asymp\frac{u'\left(1+c_{1}u\right)}{1+u},\\
						\frac{1}{b^2}	\bh_{g_{k}}^{\H}\bB_{g}\bSigma_{g}\bB_{g}^{\H}\bz_{g_{k}}	&\asymp\frac{-c_{2}uu'}{1+u},
		\end{align}
		where $ u'\!=\!\frac{1}{\bar{b}}\!\tr(\bP_{g_{[k]}}\hat{\bar{\bH}}_{g_{[k]}}^{\H}\bB_{g}\bSigma_{g_{[k]}}\bar{\bR}_{g}\bSigma_{g_{[k]}}\bB_{g}^{\H}\hat{\bar{\bH}}_{g_{[k]}})$.
		Hence, we have almost surely that
		\begin{align}
			&\bh_{g_{k}}^{\H}\bB_{g} \bSigma_{g}\bB_{g}^{\H}\hat{\bar{\bH}}_{g_{[k]}}\bP_{g_{[k]}}\hat{\bar{\bH}}_{g_{[k]}}^{\H}\bB_{g}\bSigma_{g}\bB_{g}\bh_{g_{k}}\nn\\
			&\asymp \frac{1}{\bar{b}}	\frac{u'\left(1+c_{1}u\right)}{1+u}-\frac{1}{\bar{b}}\frac{c_{0}\left(1+c_{1}u\right)^{2}-c_{1}c_{2}^{2}u^{2}-2c_{2}^{2}u}{\left(1+u\right)^{2}}uu'\label{SGI3}\\
			&=\frac{1}{\bar{b}}\frac{1-\tau_{g}^{2}\left(1-\left(1+u\right)^{2}\right)}{\left(1+u\right)^{2}}u',\label{SGI4}
		\end{align}
		where \eqref{SGI3} reduces to \eqref{SGI4}  after proper substitutions. Based on the rank perturbation lemma \cite[Lem. 3]{Hoydis2013}, we have
		\begin{align}
			u&\asymp \frac{1}{\bar{b}}\tr\left(\bar{\bR}_{g}\bSigma_{g}\right).
		\end{align}
		Also, we denote $ \frac{1}{\bar{b}}u'\asymp \bar{\bY}_{gg} $, where $  \bar{\bY}_{gg}=\frac{1}{b^{2}}\tr\left(\bP_{g_{[k]}}\hat{\bar{\bH}}_{g_{[k]}}^{\H}\bB_{g}\bSigma_{g_{[k]}}\bar{\bR}_{g}\bSigma_{g_{[k]}}\bB_{g}^{\H}\hat{\bar{\bH}}_{g_{[k]}}\right) $, which can be written as
		\begin{align}
			\bar{\bY}_{gg}\asymp \frac{1}{\bar{b}}	\sum_{j\ne k}^{\bar{K}}p_{g_{j}}\frac{\frac{1}{\bar{b}}\tr\left(\bar{\bR}_{g}\bSigma_{g}\bar{\bR}_{g}\bSigma_{g}\right)}{\left(1+\frac{1}{\bar{b}}\tr\left(\bar{\bR}_{g}\bSigma_{g}\right)\right)^{2}}   \label{t3}            \end{align}
		after applying \cite[Lems. 1, 4, 3]{Hoydis2013}.  Note that 
		\begin{align}
		&	\frac{1}{\bar{b}}\tr\left(\bar{\bR}_{g}\bSigma_{g}\bar{\bR}_{g}\bSigma_{g}\right)
			=\frac{1}{\bar{b}}\tr\left(\bar{\bR}_{g}^{-1/2}\bar{\bR}_{g}\bar{\bR}_{g}^{-1/2}\bSigma_{g}^{2}\right)\nn\\
			&=\dv{}{z}\frac{1}{\bar{b}}\tr\left(\bar{\bR}_{g}\left(\frac{1}{\bar{b}}\bB_{g}^{\H}\hat{\bar{\bH}}_{g}\hat{\bar{\bH}}_{g}\bB_{g}+\al\Id_{\bar{b}}-z\bar{\bR}_{g}\right)^{-1}\right)_{z=0}\nn\\
			&\asymp \frac{1}{\bar{b}}\tr\left(\bar{\bR}_{g}\bar{\bT}_{g}'(0)\right),\label{t2}
		\end{align}
		where
		\begin{align}
			\bar{\bT}_{g}(z)&=\left(\frac{\bar{K}}{\bar{b}}\frac{\bar{\bR}_{g}}{1+	\bar{\delta}_{g}}+\al\Id_{\bar{b}}-z \bar{\bR}_{g}\right)^{-1}\\
			\bar{\bT}_{g}'(0)&=\bar{\bT}_{g}(z)\left(\frac{\bar{K}}{\bar{b}}\frac{\bar{\bR}_{g}	\bar{\delta}_{g}'}{\left(1+	\bar{\delta}_{g}\right)^{2}}+\bar{\bR}_{g}\right)	\bar{\bT}_{g}(z)\label{t1}
		\end{align}
		with $ \bar{\bT}_{g}'(0) $ being the derivative $ \dv{	\bar{\bT}_{g}(z)}{z} $ at $ z=0 $. Use of \eqref{t1} into \eqref{t2} and substitution into \eqref{t3} gives $ \bar{\bY}_{gg} $.
		
		The derivation of  $ \mathrm{IGI}_{g_{k}} $ follows similar lines with  $ \mathrm{SGI}_{g_{k}} $. In this case, $ \bar{\bY}_{gl} $ is obtained as given in Theorem \ref{theorem:PGP}. Having obtained all terms in the SINR, the proof is concluded.
		\section{Proof of Proposition~\ref{Prop:groupPower}}\label{Prop1}

		The  power optimization problem for  group $ g $ is formulated as
		\begin{align}\begin{split}
				(\mathcal{P}1)~~~~~~~\max_{p_{g_{k}}\ge 0} ~~~	&	\sum_{k=1}^{\bar{K}}\log_{2}(1 + p_{g_{k}}	\nu_{g_{k},\mathrm{PGP}})\\
				\mathrm{s.t}~~~&\sum_{k=1}^{\bar{K}}p_{g_{k}}\le P_{g},
			\end{split}\label{MaximizationP} 
		\end{align}
		where $\nu_{g_{k},\mathrm{PGP}}= \bar{\gamma}_{g_{k},\mathrm{PGP}}/p_{g_{k}} $ does not depend on UE $ k $. The solution to $ (\mathcal{P}1) $ is obtained by the water-filling algorithm as
		\begin{align}
			p_{g_{k}}=\bigg[\mu- \frac{1}{\nu_{g_{k},\mathrm{PGP}}}\bigg]^{+},
		\end{align}
		where $ \mu $ is a parameter defining the water level to satisfy the constraint $ \sum_{k=1}^{\bar{K}}p_{g_{k}}= P_{g} $ \cite{Palomar2005}. Since $ \nu_{g_{k},\mathrm{PGP}} $ is identical for all UEs in group $ g $, the optimal powers, maximizing $ 	(\mathcal{P}1) $, are all equal and given by $ p_{g_{k}}^{\star}=P_{g}/\bar{K} $. In such case, all user rates of group $ g $ become equal to $ R^{\star}_{g} $ and the sum-rate of this group becomes $  \mathrm{\overline{SR}}_{\mathrm{PGP},g}=\bar{K} R^{\star}_{g}$.
		
		\section{Proof of Proposition~\ref{Prop:optimPhase}}\label{optimPhase}
		First, we present the following lemma, which is required in the following derivations.
		\begin{lemma}\label{lem1}
			The derivative of the trace $ \tr(\bA\bar{\bR}_{g}) $ including the covariance matrix $\bar{\bR}_{g}  $ and the matrix $ \bA $ with respect to $ \bs_{l}^{*} $, where   $ \bA $ is independent of  $ \bs_{l}^{*} $, is given by
			\begin{align}
				\pdv{ \tr(\bA\bar{\bR}_{g})}{\bs_{l}^{*}} =\diag\left( \beta_{2,g}\bH_{1}^{\H}\bA\bH_{1} \bPhi {{\bR}}_{\mathrm{IRS},k} \right).
			\end{align}
		\end{lemma}
		\proof
		We have 
		\begin{align}
			\pdv{ \tr(\bA\bar{\bR}_{g})}{\bs_{l}^{*}}&=\pdv{\tr\left( \beta_{2,g}\bA\bH_{1} \bPhi {{\bR}}_{\mathrm{IRS},k}\bPhi^{\H}\bH_{1}^{\H} \right)}{\bs_{l}^{*}}	\\
			&=\pdv{\diag\left( \beta_{2,g}\bH_{1}^{\H}\bA\bH_{1} \bPhi {{\bR}}_{\mathrm{IRS},k} \right)\bs_{l}^{*}}{\bs_{l}^{*}}\label{step2}\\
			&=\diag\left( \beta_{2,g}\bH_{1}^{\H}\bA\bH_{1} \bPhi {{\bR}}_{\mathrm{IRS},k} \right),
		\end{align}
		where \eqref{step2} is obtained by using  the property $ \tr\left(\bA \diag(\bs_{l})\right)=\left(\diag(A)\right)^{\T}\bs_{l} $.
		\endproof

		Hereafter, we denote the partial derivative with respect to $ \bs_{l}^{*} $ by $ (\cdot)' $. Starting from the  gradient of $ \bar{\gamma}_{g_{k},\mathrm{PGP}} $,  with respect to $ \bs_{l}^{*} $, which  relies on the quotient rule derivative, we obtain
		\begin{align}
		\pdv{\bar{\gamma}_{g,\mathrm{PGP}} }{\bs_{l}^{*}}=\frac{\pdv{S_{g}}{\bs_{l}^{*}}I_{g}-S_{g}\pdv{I_{g}}{\bs_{l}^{*}}}{I_{g}^{2}},\label{gam1}
	\end{align}
			where the calculation of the partial derivatives follows. In particular, $ S_{g}' $ is written as
		\begin{align}
			S_{g}'= 2p_{g_{k}}\left(1-\tau^{2}\right)\bar{\delta}_{g}\bar{\delta}_{g}', \label{sg1}
		\end{align}
		where 
		\begin{align}
			\bar{\delta}_{g}'=\frac{1}{\bar{b}}\tr\left(\bar{\bR}_{g}'\bT_{g}+\bar{\bR}_{g}\bT_{g}'\right)\label{sg2}
		\end{align}
		with $ \bT_{g}' $, being the derivative of an inverse matrix obtained by \cite[Eq. 40]{Petersen2012} as $ \bT_{g}'=-\bT_{g}(\bT_{g}^{-1})'\bT_{g} $. 
The derivative of $ \bT_{g}^{-1} $ is given by
		\begin{align}
			(\bT_{g}^{-1})'=\frac{{\bar{K}}}{\bar{b}}\frac{\bar{\bR}_{g}'\left(1+\bar{\delta}_{g}\right)-\bar{\bR}_{g}\bar{\delta}_{g}'}{(1+\bar{\delta}_{g})^{2}},\label{sg3}
		\end{align}
		where $ \bar{\bR}_{g}' $ is the derivative of $ \bar{\bR}_{g} $ with respect to $\bs_{l}^{*}  $ whose trace expression is  given by Lemma \ref{lem1}. As a result, after substituting \eqref{sg2}-\eqref{sg3} into \eqref{sg1}, we obtain $S_{g}'  $.
		
		The derivative of $ I_{g} $ is given by
		\begin{align}
			I_{g}'&=\bar{\bY}_{gg}'\left(1-\tau_{g}^{2}\left(1-\left(1+\bar{\delta}_{g}\right)^{2}\right)\right)+2\bar{\bY}_{gg}\tau_{g}^{2}\left(1+\bar{\delta}_{g}\right)\bar{\delta}_{g}'\nn\\
			&+\sum_{l\ne g}\left({\bar{\lambda}}_{l}'\bar{\bY}_{gl}+\bar{\lambda}_{l}\bar{\bY}_{gl}'\right)\frac{\left(1+\bar{\delta}_{g}\right)^{2}}{\bar{\lambda}_{g}}\nn\\
			&+\big(1+\sum_{l\ne g}\bar{\lambda}_{l}^{2}\bar{\bY}_{gl}\big)\frac{\left(1+\bar{\delta}_{g}\right)\left(2\bar{\delta}_{g}'\bar{\lambda}_{g}-\left(1+\bar{\delta}_{g}\right)\bar{\lambda}_{g}'\right)}{\bar{\lambda}_{g}^{2}},
		\end{align}
		where 
		\begin{align}
			\bar{\bY}_{gg}'& =\frac{P_{g}}{\bar{b}}\left(1-\frac{1}{\bar{K}}\right)\frac{\bar{m}_{gg}'\left(1+\bar{\delta}_{g}\right)-2\bar{m}_{gg}\bar{\delta}_{g}'}{\left(1+\bar{\delta}_{g}\right)^{3}},\nn\\
			\bar{\bY}_{gl}' &=\frac{1}{\bar{b}}\frac{P_{l}}{\bar{K}}	\frac{\bar{m}_{gl}'\left(1+\bar{\delta}_{l}\right)-2\bar{m}_{gl}\bar{\delta}_{l}'}{\left(1+\bar{\delta}_{l}\right)^{3}},\\
			\bar{\lambda}_{g}'&=\frac{\rho_{g}}{\bar{b}}P_{g}	\frac{2 m_{g}\bar{\delta}_{g}'-m_{g}'\left(1+\bar{\delta}_{g}\right)}{\left(1+\bar{\delta}_{g}\right)^{3}\bar{\Psi}_{g}^{2}}
		\end{align}
		are derived easily by applying basic derivative rules.
		For the derivatives of $ \bar{m}_{g} $, $ \bar{m}_{gg} $, and $ \bar{m}_{gl} $, we notice that that they have a similar expression. Thus, we resort to the definition of  a new function $ f_{g}\!\left(\bA\right) $, which includes them under specific values of its parameters $ g $  and $ \bA $, and we compute its derivative. Thus, by defining
		\begin{align}
			f_{g}\!\left(\bA\right) =\frac{\frac{1}{\bar{b}}\tr\left(\bar{\bR}_{g}\bT_{g}\bA\bT_{g}\right)}{1-\frac{\frac{\bar{K}}{\bar{b}}\tr\left(\bar{\bR}_{g}\bT_{g}\bar{\bR}_{g}\bT_{g}\right)}{b \left(1+\bar{\delta}_{g}\right)^{2}}},
		\end{align}
		we have $ 	\bar{m}_{g}=f_{g}\left(\bB_{g}^{\H}\bB_{g}\right) $, $ 	\bar{m}_{gg}=f_{g}\left(\bar{\bR}_{g}\right) $, and $ 	\bar{m}_{gl}=f_{l}\left(\bB_{l}^{\H}\bar{\bR}_{g}\bB_{l}\right) $. Note that the dependence on the RBM is found on $ \bar{\bR}_{g} $ while $ \bB_{g} $ appears no such dependence. Also, we define $ q(\bC)=\tr\left(\bar{\bR}_{g}\bT_{g}\bC\bT_{g}\right) $. After  some lengthy algebraic manipulations, we obtain $ f_{g}'\!\left(\bA\right) $ in \eqref{fa},
		where   $ q'\!(\bC) $ for $\bC\!=\!\bA, \bar{\bR}_{g}  $  written as $ 	q'\!(\bC)\! =\!	\tr\!\left(\bar{\bR}_{g}'\bT_{g}\bC\bT_{g}\!+\!\bar{\bR}_{g}\bT_{g}'\bC\bT_{g}\!+\!\bar{\bR}_{g}\bT_{g}\bC'\bT_{g}\!+\!\bar{\bR}_{g}\bT_{g}\bC\bT_{g}'\right)\!, $
		with  $  \bT_{g}'$ given above while the derivatives of the traces are obtained by using Lemma \ref{lem1}. Hence, after suitable substitutions in \eqref{fa}, we obtain $ 	\bar{m}_{g}' $, $ 	\bar{m}_{gg}' $, and $ 	\bar{m}_{gl}' $. 
	\end{appendices}
	
	\bibliographystyle{IEEEtran}

	\bibliography{mybib}
\end{document}